\documentclass[fleqn,usenatbib]{mnras}

\usepackage{newtxtext,newtxmath}

\usepackage[T1]{fontenc}

\DeclareRobustCommand{\VAN}[3]{#2}
\let\VANthebibliography\thebibliography
\def\thebibliography{\DeclareRobustCommand{\VAN}[3]{##3}\VANthebibliography}

\usepackage{graphicx}	
\usepackage{amsmath}	
\usepackage[dvipsnames]{xcolor}

\title[Microlensing BH Shadows]{Microlensing Black Hole Shadows}

\author[H. Verma et al.]{
Himanshu Verma$^{1}$\thanks{E-mail: verma.himanshu002@gmail.com}
and Joseph Silk$^{2,3,4}$\thanks{E-mail: silk@iap.fr}
\\
$^{1}$Department of Physics, Indian Institute of Technology Bombay, Powai, Mumbai, Maharashtra, 400076, India\\
$^{2}$Institut d'Astrophysique de Paris (UMR7095: CNRS \& UPMC- Sorbonne Universities), F-75014, Paris, France\\
$^{3}$Department of Physics and Astronomy, The Johns Hopkins University Homewood Campus, Baltimore, MD 21218, USA\\
$^{4}$BIPAC, Department of Physics, University of Oxford, Keble Road, Oxford OX1 3RH, UK
}

\date{Accepted XXX. Received YYY; in original form ZZZ}

\pubyear{2023}

\begin{document}
\label{firstpage}
\pagerange{\pageref{firstpage}--\pageref{lastpage}}
\maketitle

\begin{abstract}
A detailed analysis is presented of gravitational microlensing by intervening compact objects of the black hole shadows imaged by the Event Horizon Telescope (EHT). We show how the center, size, and shape of the shadow depend on the Einstein angle relative to the true/unlensed shadow size, and how the location of the lens affects the shift, size, and asymmetry of the black hole shadow due to microlensing. Assuming a supermassive black hole (SMBH) casts a circular-shaped true shadow, microlensing can create an asymmetry of up to approximately 8\%, which is twice the asymmetry caused by the SMBH's spin and its tilt relative to us. Furthermore, the size can be enhanced by $\sim$50\% of the true shadow. Currently, the terrestrial baselines of EHT lack the resolution to detect microlensing signatures in the shadows. However, future expansions of EHT, including space-based baselines at the Moon and  L$_2$, could potentially enable the detection of microlensing events. For Sgr~A$^*$, an event rate of 0.0014 per year makes the microlensing phenomena difficult to observe even with space-based baselines for the stellar population in the stellar bulge and stellar disk for lens masses $\sim M_\odot$. Nonetheless, the presence of a cluster of 20,000 stellar-mass black holes in the central parsec of the Milky Way, expected to arise from dynamical friction acting on  infalling stellar clusters, significantly boosts the event rate. Hence, continuously monitoring the shadow of Sgr~A$^*$ could offer novel insights into the compact object population surrounding the Galactic Center.
\end{abstract}




\section{Introduction}

\label{sec:Intro}
Black holes are natural laboratories for studying the most extreme environments found in strong gravity. Event Horizon Telescope (EHT) exploits the Very Large Baseline Interferometry (VLBI) facilities across the globe, forming an effective baseline of earth size and resulting in a resolution of $\sim 25~\mu$as at 1.3 mm wavelength. One of the most astonishing discoveries of EHT has been the \textit{direct} imaging of the shadows of the supermassive black holes (SMBH) at the center of the Messier 87 galaxy (M87$^*$~\citep{EventHorizonTelescope:2019dse, EventHorizonTelescope:2019uob, EventHorizonTelescope:2019jan, EventHorizonTelescope:2019ths, EventHorizonTelescope:2019pgp, EventHorizonTelescope:2019ggy}) and of the Milky Way (Sagittarius~A$^*$ (Sgr~A$^*$)~\citep{2022ApJ...930L..12E, 2022ApJ...930L..13E, 2022ApJ...930L..14E}). 

The theory that a black hole in the backlight of a background uniform source at infinity would cast a shadow of its event horizon was first advanced by Bardeen~\citep{1974IAUS...64..132B}. Subsequently, \citep{1979A&A....75..228L} and ~\citep{Falcke:1999pj, Falcke:2013ola} developed a more realistic shadow model by considering the emission from the accretion disc surrounding the black hole. The average angular radius of the shadow for a rotating black hole is widely understood to be (5$\pm$0.2)$GM/(D_sc^2)$, where $D_s$ is the source's distance from the earth and $M$ is the SMBH's mass (see eg.~\citep{2000CQGra..17..123D, 2004ApJ...611..996T, 2006PhRvD..74f3001B, 2010ApJ...718..446J, Psaltis:2014mca, Johannsen:2016uoh}). The standard deviation about the average radius arises from the asymmetric shape of the shadow, which can occur via changes in spin value and spin axis tilt (for a review of the precise dependence, see~\citep{Cunha:2018acu, Gralla:2019xty, Perlick:2021aok}). It is of the utmost importance to measure the size and shape of the shadow precisely because, assuming the Kerr metric describes the spacetime around a spinning black hole, it offers a unique probe for determining the spin and mass of the core SMBH. Furthermore, when the mass and spin of the SMBH are precisely determined through other observational methods, such as pulsars in proximity to the SMBH~\citep{Liu:2014uka}, stellar orbits encircling the SMBH~\citep{2019Sci...365..664D}, and the electromagnetic spectrum from accreting gas around the SMBH~\citep{2017RvMP...89b5001B}, the observed characteristics of the shadow will offer a chance of  identifying potential deviations from the Kerr hypothesis. This presents an opportunity to assess and test modified theories of general relativity.

Another \textit{indirect} technique for measuring the shadows involves photometric microlensing, where gravitational bending around massive foreground objects leads to an apparent brightening in the background source~\citep{1964MNRAS.128..295R, 1989AJ.....98.1989I}. This technique utilizes the microlensing phenomena induced by stars and compact objects within a galaxy that strongly lens a quasar, causing time-varying brightening~\citep{1986A&A...166...36K, 1987A&A...171...49S, 2004ApJ...605...58K, 2010GReGr..42.2127S}. It has been used to study accretion disks around the quasars~\citep{1988A&A...194...54G, 1997ApJ...483L..13G, 1999ApJ...524...49A, 1999MNRAS.302...68F, 2005ApJ...628..594M, 2008ApJ...673...34P} and it has been proposed as a tool for imaging the vicinity of the SMBHs in quasars. Although the quasar's SMBH silhouette itself is too small for direct resolution, its accretion disk with a shadow creates a distinctive signal in the quasar's light curve~\citep{2012MNRAS.423..676A, 2015ApJ...814L..26M}. This technique has also been extended to self-lensing in binary SMBH systems~\citep{DOrazio:2017ssb, DOrazio:2019ens, Ingram:2021gar}. Furthermore, 
their shadows may self-lens due to strong gravitational effects and hence produce a dip in the light-curve~\citep{Davelaar:2021eoi, Davelaar:2021gxx}, and a similar study has also been conducted in a stellar-mass black hole with binary companion star~\citep{Gott:2018ocn}. These scenarios of indirect shadow imaging hold considerable promise for future observational endeavors~\citep{2013CQGra..30x4003F}.

Here, we present the formalism for the microlensing of the directly observable shadow. We apply the framework of microlensing on each point source of the boundary of the shadow due to a point lens in the foreground. Varying angular separations of the boundary sources from the lens lead to the apparent brightening and angular shift of each source on the boundary. This will be manifest as an overall distortion in the shadow. We will call the distorted shadow the \textit{microlensed shadow}. In the case of relative motion between the source and the lens, the proposed distortion in the shadow will be a time-varying phenomenon.

We present an analytic dependence of the microlensed shadow's shape on the lens size and its variation with lens angular separation, assuming a circular shape of the true shadow. The primary signature of microlensing on the observed shadow is to enhance the size of the shadow and make the shadow asymmetric. The maximum enlargement in the size of the shadow turns out to be 50\% of the true size of the shadow. The highest asymmetry, however, can reach up to $\sim$8\% of the average radius of the microlensed shadow, which is twice the maximum asymmetry (4\%~\citep{Psaltis:2014mca}) arising due to inclination and the spin of the SMBH. As the non-zero spin of the SMBH may also cause asymmetry in the true shadow, a non-zero spin may contaminate the effect of microlensing. In this work, we restrict our analysis to the circular shape of the true shadow. We demonstrate the formalism of calculating microlensed shadow to the case of Sgr~$A^*$ for which the abundances of stellar lenses in the Milky Way
are well constrained observationally. We discuss the detectability assuming the achievable resolution from various baseline configurations such as Earth-size baselines (already achieved by current EHT observations) and futuristic Earth-space baselines~\citep{Roelofs:2019nmh, 2020AdSpR..65..821F, Mikheeva:2020bqj, Gurvits:2022wgm, Tiede:2022grp, Chael:2022meh, 2022AcAau.196...29R}. We compute the distribution of event rates throughout the microlensing event durations by employing an analytic stellar model of the Milky Way. Our analysis indicates that the total event rate cannot exceed $10^{-2}$ per year. Nevertheless, we discuss scenarios where the event rate might experience enhancements.

In a broader context, the ongoing microlensing event during the observation of the shadow can mimic the deviation in the size and shape of the shadow from the null hypothesis of the Kerr background metric, which is also predicted by studies on searching beyond standard phenomena, such as testing the no-hair theorem of black holes~\citep{Johannsen:2010xs, Johannsen:2010ru}, searching for super-spinning black holes~\citep{Bambi:2008jg, Bambi:2019tjh}, testing the Kerr hypothesis~\citep{Bambi:2011yz, Bambi:2015rda} and probing light-scalar hair around the SMBH~\citep{Cunha:2015yba, Cunha:2019ikd}. Therefore, before interpreting the deviation of a shadow from the null hypothesis in search of the beyond-standard phenomenon, one should rule out the standard possibility of a foreground compact object microlensing the shadow. Hence, the microlensing of a black hole shadow should be treated as a standard astrophysical background phenomenon for the search of beyond standard physics using the shadow, which can be reduced by proper modeling of the microlensing event rate distribution over the event durations discussed in this paper.

In sec.~\ref{sec:MicrolensingShadow}, we briefly review the microlensing of a point source due to a point lens. We then develop the formalism for microlensing of the shadow and give a generic expression for the distorted shadow caused by microlensing. In sec.~\ref{sec:Char}, we characterize the microlensing signal of the shift in the shadow center and the radial variation for the microlensed shadow with position angle. The enlarged shadow will be described in sec.~\ref{subsec:size}, and the induced asymmetry will be covered in sec.~\ref{subsec:asymmetry}. Finally, in sec.~\ref{sec:detectability}, we discuss the detectability of the microlensed shadow of Sgr A$^*$ in an EHT extension with baselines at Moon and L$_2$ positions. The anticipated duration of the events, optical depth, and the distribution of event rate over the event durations are calculated assuming a stellar population for the bulge and the disk of the Milky Way. The potential ways to enhance the event rate are also discussed. We then summarize our findings in sec.~\ref{sec:summary}.

\section{Microlensing of a black hole shadow}
\label{sec:MicrolensingShadow}
In this section, we will study the distortion of a black hole shadow due to the microlensing effect of a lens anywhere between  Sgr~A$^*$ and us. The shadow of Sgr~A$^*$ is formed at the center of the Milky Way, where we have a huge stellar density in the central nuclear star cluster according to ref.~\citep{2018A&A...609A..26G, 2018A&A...609A..27S, 2018A&A...609A..28B}. For example, the total luminous stellar mass within 0.01 pc (1 arcsec) from Sgr~A$^*$ is 180$\pm$30 $M_\odot$. Furthermore, the stellar bulge population~\citep{2016A&A...587L...6V} of mass $2\times10^{10}~M_\odot$ within 1~kpc may also perturb the shadow. Hence, we can expect a significant perturbation of the shadow from the stellar components. Furthermore, the possibility of an orbiting IMBH~\citep{Deme:2019zyv} and stellar-mass black hole cluster in the central parsec of the Milky Way~\citep{Miralda-Escude:2000kqv, 2018Natur.556...70H} may also distort the shadow. The actual effect would depend upon the mass of the lenses as well. We will study the effect of lenses with mass ranging from planetary mass $M \sim 10^{-6}~M_\odot$ and up to lenses of mass $M \sim 10^{4}~M_\odot.$ 

\subsection{Black hole shadow and lensing ring}
The shadow is one of the most distinguishing features of the direct image of an SMBH with a surrounding accretion disk. The photons from the accretion disk undergo strong gravitational lensing, which causes the shadow to form. Therefore, the ``observed shadow'' extends up to the lensed position of the inner edge of the accretion disk~\citep{1979A&A....75..228L, Falcke:1999pj,2010ApJ...718..446J, Falcke:2013ola}. 

Apart from the shadow, another feature of the black hole image is the lensed image of the optically thin accretion disk, which is yet to be observed by EHT or EHT-like VLBI facilities. The existence of the null geodesics surrounding black holes, whose order can be identified by the number of half revolutions of the photons around the black hole before escaping, causes the lensed images to develop. We will call all these lensed images ``lensing rings'' whereas some refer to all lensing rings with half rotations greater than two as photon rings. In principle, an infinite number of lensing rings exist, but all are formed around the critical curve (Bardeen's shadow). Their thickness becomes sharper, and they form nearer to  Bardeen's shadow as the order of the lensing ring rises~\citep{2010GReGr..42.2269B, Johnson:2019ljv, Gralla:2020yvo}. As a result, the size and shape of the lensing rings approach  Bardeen's shadow. They are thus asymptotically wholly governed by the characteristics of the black hole alone. Therefore, investigating the impact of the microlensing signal on Bardeen's shadow is a natural choice that will not be affected by the accretion physics\footnote{In principle, the shape of the lensing ring does not depend upon the accretion physics, their observed shape may depend on it.  This is because the accretion disk may have an inherent variability, and moving hot spot features, which can make the observed black hole image asymmetric~\citep{EventHorizonTelescope:2019pgp, 2019ApJ...870....6Z} and hence their lensing rings~\citep{2022A&A...668A..11P}.}. However, this paper's formalism of microlensing of a shadow is equally valid for both the observed shadow and the lensing rings. We shall use the term \textit{boundary of the shadow} to refer to any cases of the boundary of Bardeen's shadow, the boundary of the observed shadow, and the boundary of the lensing rings without losing the generality of the effect. We will also assume a circular shape boundary of the true shadow.

We will first review the microlensing of a point source due to a point lens. We then develop the theory of the effect of microlensing on the shadow.

\subsection{Brief review of microlensing}
\label{subsec:MLreview}
\begin{figure}
    \centering
    \includegraphics[width=\columnwidth]{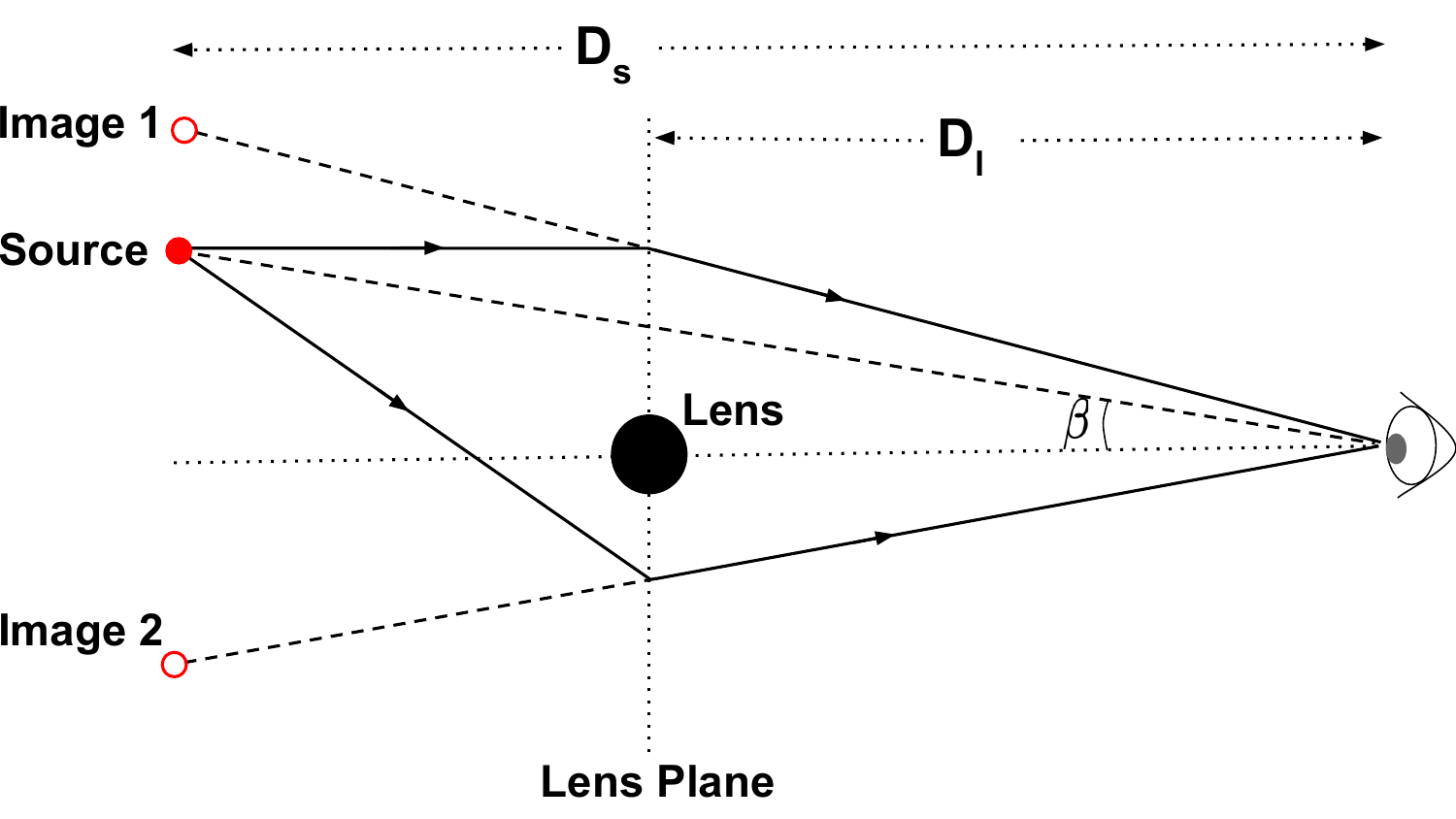}
    \caption{Ray diagram of gravitational lensing of a point source due to a point lens. The light rays coming from the source bend around the curved space around the lens (black-blob) and forms two apparent lensed images (Image 1 and Image 2) of the source.}
    \label{fig:MicrolensingCartoon}
\end{figure}
Consider a point source in the sky at $D_s$ distance away from us as shown in fig.~\ref{fig:MicrolensingCartoon}. A point lens of mass $M$ is also present between the source and us. Hence, its distance $D_l$ from us is always less than $D_s$. Due to the gravitational lensing effect, the light rays from the source would bend around the lens, giving the illusion of two separate point images of the source. To comprehend the geometry governing the production of the images, we imagine a lens plane that travels through the lens and is perpendicular to the line connecting the lens and ourselves. We can use the line between the source and us to project the source onto the lens plane. An angular separation vector $\vv{\beta}$ of the source with respect to the lens can then be defined such that $D_l \vv{\beta}$ represents the position vector of the projection of the source on the lens plane. The angular position vector of the two images can be ascertained by
\begin{eqnarray}
\label{eq:ImagePos}
\vv{\theta}_\pm &=& \left(1 \pm \sqrt{1 + 4\frac{\theta_E^2}{\beta^2}}\right)\frac{\vv{\beta}}{2},
\end{eqnarray}
where $\theta_E$ is the Einstein angle, and determines the angular scale of the separation between the two images and its dependence on the parameters of the source and the lens, scaling as~\citep{2012ARA&A..50..411G}
\begin{eqnarray}
    \theta_E &\approx& 2.85 \text{ mas }\sqrt{ \frac{M}{1 \textrm{ M}_{\odot}}\frac{1 \textrm{ kpc}}{D_s}\left(\frac{D_s}{D_l} - 1\right)}.
\end{eqnarray}
According to eq.~\ref{eq:ImagePos}, the angular position vector with respect to the lens is along $\vv{\beta}$ for one image and opposite to $\vv{\beta}$ for the other. As a result, the two images will always be aligned on the lens plane along the line connecting the lens and the projected source. There exists an imaginary ring of angular radius $\theta_E$ centered at the lens called an Einstein ring such that one of the images would be outside of the ring, and the second image would be inside the ring. 

Another consequence of lensing is that the apparent brightness of the two images gets magnified as compared to the true brightness of the point source. The respective magnification in the brightness for the two images is given by,
\begin{eqnarray}
\label{eq:ImageMagnifications}
\mu_\pm &=&  \frac{\beta^2 + 2\theta_E^2}{2\beta\sqrt{\beta^2 + 4 \theta_E^2}} \pm \frac{1}{2}.
\end{eqnarray}
More details of the above discussion can be found in~\citep{1992grle.book.....S, Narayan:1996ba, Wambsganss:1998gg}.

\begin{figure}
	\centering
	\includegraphics[width=\linewidth]{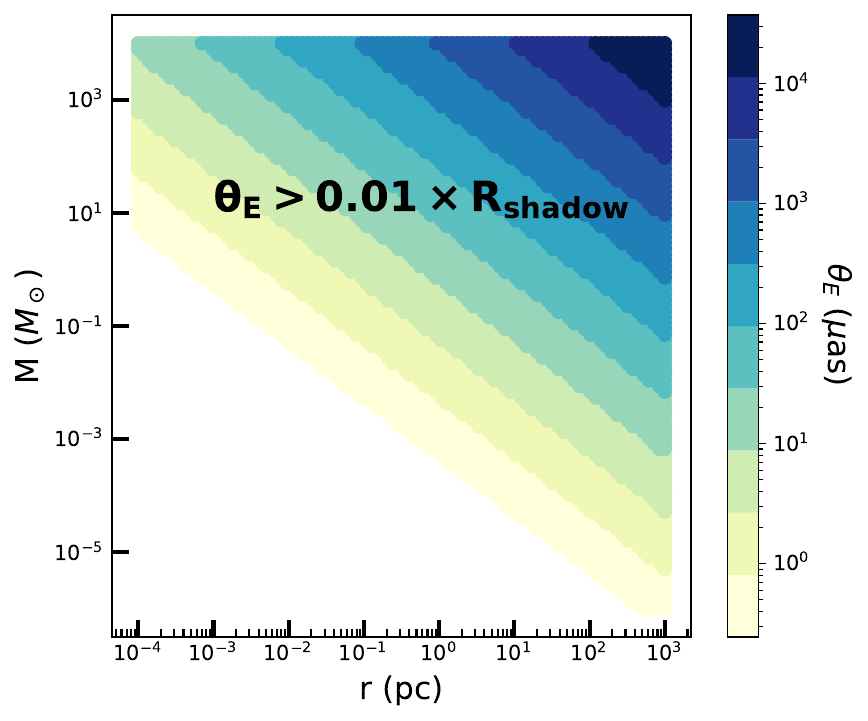}
	\caption{This plot depicts the relevant parameter space of mass $M$ and radial distance $r$ of the lens from Sgr~A$^*$ towards Earth. Parameters in the colored region can potentially induce measurable distortion in the shadow expected to be $\sim \theta_E$, where the corresponding Einstein angle can be read from the color bar. Assuming 1\% accuracy in the true radius $R_\textrm{shadow} = R$ of the shadow, we obtain the lower cut-off in the parameter space by $\theta_E = 0.01 R$ i.e. $M \approx 6\times10^{-8}~M_\odot (8200~\textrm{pc}/r - 1)$ for Sgr~A$^*$. This parameter region will be further refined later in the paper after quantifying the exact distortion due to microlensing. }
	\label{fig:ParamSpace}
\end{figure}

If the separation between the two images is small compared to a telescope's resolution, then the two distinct images will not be observable. However, the telescope would see a magnification-weighted average image called the \textit{centroid}. The angular position of such a centroid relative to the lens will be~\citep{1995A&A...294..287H, 1995ApJ...453...37W, Gould:1996nb, 2000ApJ...534..213D, 2017IJMPD..2641015N}
\begin{eqnarray}
\label{eq:pos_c}
\vv{\theta}_c &\equiv& \frac{\vv{\theta}_+ |\mu_+| + \vv{\theta}_-|\mu_-|}{|\mu_+| + |\mu_-|} =  \frac{\beta^2 +3\theta_E^2}{\beta^2 +2\theta_E^2}\vv{\beta}.
\end{eqnarray}
The centroid always lies on the line joining the lens to the projected source on the lens plane. The astrometric angular shift in the centroid relative to the true angular position of the source is quantified to be
\begin{eqnarray}
\label{eq:shift_c}
\vv{\delta}_c &\equiv& \vv{\theta}_c - \vv{\beta}=  \frac{\theta_E^2}{\beta^2 +2\theta_E^2}\vv{\beta},
\end{eqnarray}
which is an angular vector defined such that $D_l \vv{\delta}_c$ is the shift of the centroid on the lens plane from the true position of the projected source. The photometric magnification in the flux of the centroid would be,
\begin{eqnarray}
\label{eq:amp_c}
\mu_c &\equiv& |\mu_+| + \mu_-|  =\frac{\beta^2 + 2\theta_E^2}{\beta\sqrt{\beta^2 + 4 \theta_E^2}}.
\end{eqnarray}

In the case of a relative proper motion of the source relative to the lens, we expect $\vv{\beta}$ to be time-varying; therefore, the angular position of the centroid would change with time. The phenomenon of variation of this apparent position of the source with time is called \textit{astrometric microlensing}. On the other hand, a time variation in the magnification of the centroid is called \textit{photometric microlensing}. 

Given that the shift in the centroid of a point source is $\sim \theta_E$, we can expect the distortion on a shadow of angular radius $R$ would also be of the order of the Einstein angle. Therefore, assuming 1\% accuracy in the measurement of the shadow, in fig.~\ref{fig:ParamSpace}, we extract a target parameter space of the lens for which the Einstein angle and hence the distortion could be larger than 1\%.

\subsection{Theory of the microlensed shadow}
\label{subsec:MLshadow}
We will now summarize our formalism for describing the effect of microlensing on the boundary of a black hole shadow. We expect massive objects such as stars, stellar black holes, neutron stars, white dwarfs, brown dwarfs,  dark matter, and even primordial black holes to surround a supermassive black hole at the center of a galaxy. Furthermore, these massive objects could also be floating between us and the SMBH. If these objects are close enough to the shadow along our line of sight during the observational period of the EHT-like telescope, they might serve as lenses for each point source of the boundary of the shadow. We expect that light rays originating from different points of the boundary of the shadow would get deflected due to gravitational lensing by intervening objects. 

To understand the effect of lensing on the boundary of the shadow, we assume that the true boundary of the shadow is circular. We also assume the boundary to be a collection of point sources. Although the influence of the lenses on each point source of the boundary will undoubtedly produce two images separated ($\sim \theta_E$) from one another, it would be difficult for EHT-like VLBI facilities to resolve such effects. Hence we make an assumption that only the centroid of each point source would be observable by the telescope. We apply the formalism of microlensing described in sec.~\ref{subsec:MLreview} to each point source that makes up the boundary of the shadow.

\begin{figure}
	\centering
	\includegraphics[width=\linewidth]{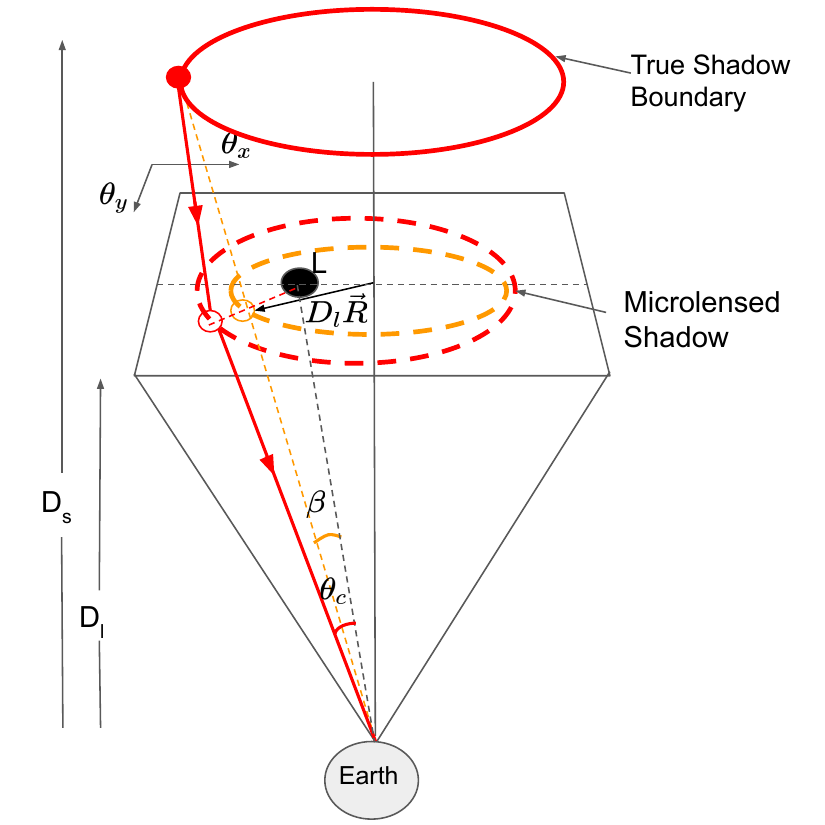}
	\caption{This schematic cartoon illustrates the anticipated outcome from the microlensing of the shadow. The red-solid ring shows the true boundary of the shadow, which may be thought of as a collection of several point sources. The orange-dashed circle indicates the boundary projection onto the lens plane. As depicted in the cartoon, each point source on the boundary will be at a different angular separation from the lens L. Hence, it is expected that the shift in the centroid (given by eq.~\ref{eq:shift_c}) will differ for various point sources across the boundary. The centroid of each point source on the boundary can then be located to determine the microlensed shadow. The expectation is that the microlensed shadow would be a closed contour of center, size, and shape different from the true boundary of the shadow. A cartoon of the microlensed shadow is shown with a red-dashed contour, which is distorted compared to the true shadow.}
	\label{fig:LensedRingcartoon}
\end{figure}

In fig.~\ref{fig:LensedRingcartoon}, we depict a schematic diagram of the true boundary of the black hole shadow, represented by a red-solid circle. The foreground lens $L$ is indicated by the black-filled blob on the lens plane. The true boundary is also projected on the same plane. The projected boundary of the shadow is shown as a orange-dashed curve on the lens plane. It is assumed that the lens is distance $D_l$  away from us. In contrast to the previous sec.~\ref{subsec:MLreview}, we now have numerous point sources as a part of the true boundary at the same distance $D_s$. However, the angular separation vector $\vv{\beta}$ of the point sources on the projected boundary from the lens would be different. Due to the small angular scales involved, we can define a Cartesian coordinate on the lens plane with the center of the projected shadow as the origin. The x-axis is assumed to pass through the center of the projected shadow and parallel to the motion of the lens. Each point source is identified on the projected boundary by a position angle $\phi$ measured with the x-axis. The shift in the centroid due to the microlensing will not be the same for all the point sources because of their different angular separation from the lens. As a result, we can anticipate a distorted boundary of the shadow. The red-dashed curve shows the distorted shadow in fig.~\ref{fig:LensedRingcartoon}, which we will refer to as the \textit{microlensed shadow}.

\begin{figure*}
	\centering
	\includegraphics[width=\textwidth]{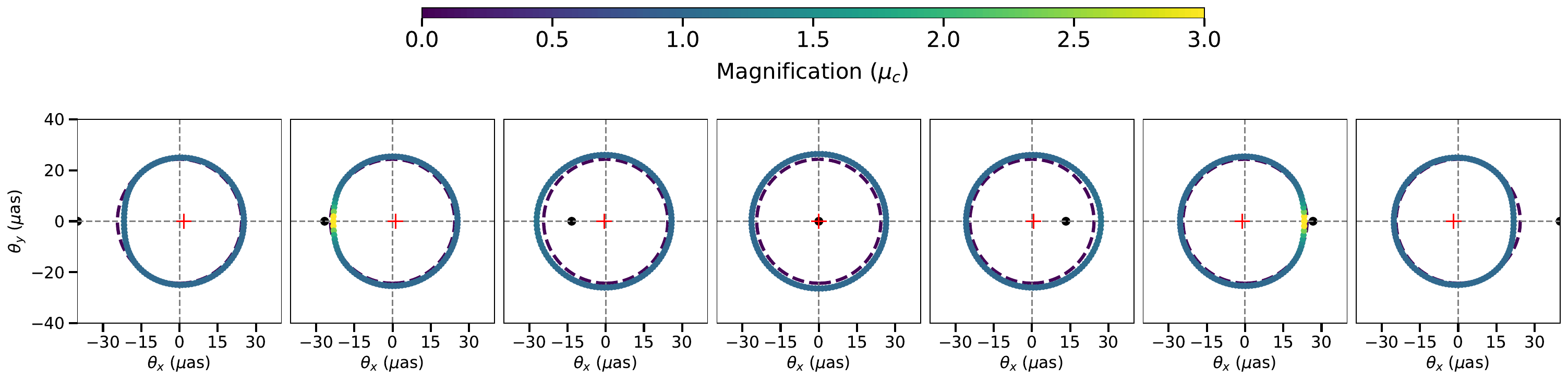}
	\caption{As an illustration of the phenomenon of microlensing of the black hole shadow, we have assumed the true shape of the boundary of the shadow of Sgr~A$^*$ ($D_s = 8.2$ kpc) to be a perfect circle of radius $R = 24.35~\mu$as. The blue-dashed circle represents the true boundary of the black hole shadow. A lens is indicated by a black dot with an Einstein angle $\theta_E = 7.8~\mu$as. This Einstein angle corresponds to the mass of the lens to be $M = 1~M_\odot$ and distance $r = 0.5$ pc from Sgr~A$^*$. The microlensed images of an arbitrary source on the true boundary of the shadow trace the dotted contour which is the final microlensed shadow under centroid approximation (see sec.~\ref{subsubsec:CentroidApproximation} for the validity). The color of the dots represents the photometric magnification of the images, where the magnification value can be read from the color bar. The shifted center of the apparent shadow is shown with a red marker. Various panels show the variation of the microlensed shadow as the lens passes in the foreground of the true shadow.}
	\label{fig:lensedshadow}
\end{figure*}

We define the true boundary of the shadow to be a circular ring with an angular radius $R$ to calculate the impact of microlensing on the shadow. The angular position vector of a point source on the boundary of the ring can be labeled by
\begin{eqnarray}
\vv{R} &\equiv& \begin{bmatrix}
R\cos{\phi}\\
R \sin{\phi}
\end{bmatrix},
\end{eqnarray}
where $\phi\in[0,2\pi)$. Additionally, the angular position vector of the lens $\vv{\xi}$ is defined such that $D_l \vv{\xi}$ is the position vector of the lens on the lens plane from the origin. As a result, the angular position vector of a point source on the true boundary relative to the lens would then be determined by
\begin{eqnarray}
\label{eq:betavec_ring}
\vv{\beta} &=& \vv{R} - \vv{\xi},
\end{eqnarray}
and the magnitude of the angular separation between the lens and any point source on the boundary of the shadow is given by the following formula,
\begin{eqnarray}
\label{eq:beta_ring}
\beta &=& \sqrt{\xi^2 +R^2 - 2 \xi R \cos{\phi}}.
\end{eqnarray}
Given the position of the point sources parameterized by $\phi$ on the boundary relative to the lens by eq.~\ref{eq:betavec_ring}, we can obtain the apparent lensed location of the sources or centroids using  eq.~\ref{eq:pos_c}. The angular positions of these centroids expressed in terms of the position angle $\phi$ are as follows,
\begin{eqnarray}
    \label{eq:thc(phi)}
    \vv{\theta}_c(\phi) = \frac{\beta(\phi)^2 + 3\theta_E^2}{\beta(\phi)^2 + 2\theta_E^2}\vv{\beta}(\phi).
\end{eqnarray}

The above expression of the centroid is an angular position vector from the lens location since eq.~\ref{eq:betavec_ring} provides the angular position of the centroid, which assumes the lens to be at the origin. But to obtain the microlensed shadow, the origin (denoted by $O$) has been chosen at the center of the true shadow. Hence, the locations of the centroids $I_c$ with the true center as the origin are determined as
\begin{eqnarray}
\vv{OI}_c(\phi) &=& \vv{\theta}_c + \vv{\xi}= \vv{R} + \frac{\theta_E^2}{\beta(\phi)^2 + 2\theta_E^2}\vv{\beta}(\phi).
\end{eqnarray}
To obtain the expression of the contour of the microlensed shadow, we must first determine the new center of the apparent microlensed shadow labeled by $C$. We define the shifted geometrical center of the microlensed shadow as
\begin{eqnarray}
	\label{eq:lcenter}
    \vv{OC} &\equiv& \frac{\vv{OI}_c(\pi) + \vv{OI}_c(0)}{2}.
\end{eqnarray}
We will call this the  \textit{lensed center}. The lensed center of the microlensed shadow will always align with the lens and the center of the true shadow. A detailed dependence of the lensed center on the lens and true shadow characteristics will be discussed in sec.~\ref{sec:Char}. 

Now, to quantify the shape of the contour of the centroids with regard to the boundary of the microlensed shadow, we introduce an angular radial vector $\vv{R}_L (\phi)$. It starts at the lensed-center and ends at the centroid of a point source on the true boundary. The expression of $\vv{R}_L (\phi)$ originated from the shifted center $C$ is given by,
\begin{eqnarray}
\label{eq:RL}
\vv{R}_L = \vv{OI}_c - \vv{OC}= \vv{R}- \vv{OC} + \frac{\theta_E^2}{\beta(\phi)^2 + 2\theta_E^2}\vv{\beta}(\phi).
\end{eqnarray}
It should be noticed that the position angle of $\vv{R}_L(\phi)$ will not always be the same as the position angle ($\phi$) of $\vv{R} $. This suggests that the microlensing will also cause a twist along the boundary. The angle, say $\phi_L$, formed by $\vv{R}_L(\phi)$ with the positive x-axis can be found by using the components of $\vv{R}_L(\phi)$ along the x-axis and the y-axis as indicated by $R_{L,x}$ and $R_{L,y}$, respectively. Hence, the twisted position angle is obtained by,
\begin{eqnarray}
\label{eq:phiL}
\phi_L &=& \tan^{-1}\left(\frac{R_{L,y}}{R_{L,x}}\right).
\end{eqnarray}

Another consequence of microlensing on the shadow would be non-uniform brightness magnification for the various portions of the boundary. The magnification in the flux (see eq.~\ref{eq:amp_c}) varies for the point sources across the boundary as a result of the same fact that the angular separation $\beta$ differs for different sites on the boundary. Therefore, the non-uniform brightness of the boundary would be a sign of photometric microlensing of the shadow. A drawback of using eq.~\ref{eq:amp_c} is in the case of the lens lying on the ring. The point source perfectly aligning with the lens will have the magnification to be theoretically infinite, but practically will be limited by the total energy emitted by the point source per unit time and per unit area around the point source,  and hence remains finite. Although the expression of magnification under the point source approximation blows up for the perfect alignment, the magnification-weighted average angular position of the image (centroid) given by eq.~\ref{eq:thc(phi)}, which is used to calculate the distorted shape of the shadow, will be zero for the point source perfectly aligned with the lens and remains finite for all other values of the lens-source separation.

In fig.~\ref{fig:lensedshadow}, we demonstrate the effect of microlensing on the true boundary of the shadow shown by the dashed-blue circle having a center at $O$. To illustrate the effect, we consider Sgr~A$^*$ located at a distance $D_s = 8.2$ kpc. We assume the true black hole shadow has a circular shape with a radius of $R = 24.35~\mu$as. The figure exhibits how the microlensed shadow changes for various angular positions of the lens of mass 1 $M_\odot$ and at a distance of $r = 0.5$ pc from Sgr~A$^*$ in the foreground of the shadow. Here, we display the true boundary of the shadow, the true center, and the lens, together with the microlensed shadow and the lensed-center. These panels reveal a few general characteristics of the microlensing of the shadow. First, the shadow shifts away from the lens, which is shown by the lensed-center (see the red marker). However, when the lens and the true shadow are perfectly aligned, the lensed center also coincides with the true center. Second, the shadow's shape is warped, and its size is enlarged. The shadow takes on the shape of a circle when the lens is perfectly aligned with the true shadow. The third effect is the asymmetry in the brightness of the boundary. This is due to the uneven magnifications for various portions of the boundary of the shadow. In the case of the lens and the true shadow being perfectly aligned, the only remaining characteristics of microlensing are the shadow's enlarged size and an overall magnification in the brightness of its boundaries. However, the brightness asymmetry and the asymmetric shape vanish when the lens and the true shadow are perfectly aligned.

The figure shows that the boundary's asymmetric brightness is important when the lens is close to the true shadow. In contrast to the asymmetry in brightness, the asymmetric shape of the shadow survives with a wider angular separation of the lens from the true shadow. This behavior can be understood qualitatively using a point lens's microlensing of a point source. It is well-known that the photometric magnification signal fades away much more quickly than the astrometric shift signal at large angular separations of the lens. This is because the astrometric shift (given by eq.~\ref{eq:shift_c}) falls off as $\sim 1/\beta$  at large $\beta$, whereas the photometric magnification (provided by eq.~\ref{eq:amp_c}) falls off as $\sim 1/\beta^4$. That is why, even if the lens were far from the shadow, the strength of the astrometric distortion is substantially stronger than the photometric distortion.

The photometric asymmetry caused by microlensing could be important for the event when the lens passes very close to the boundary. It will change the morphology of the boundary, and microlensing being a time-varying phenomenon, it will add another time-scale (Einstein crossing time $t_E\equiv\theta_E/\mu$, where $\mu$ is the angular velocity of the lens relative to the source) into the observability of the boundary of the shadow apart from the innermost stable circular orbital time scale. However, if the integration time (time taken by EHT-like telescope to collect enough photons to calculate the visibility) is larger than $t_E$, the photometric effect will be washed out. Given that the astrometric distortion will have its influence from a much larger angular separation of the lens from the shadow as compared to the photometric effect, we will be focusing on the impact of astrometric microlensing on the shadow.

\subsubsection{Centroid approximation and its validity}
\label{subsubsec:CentroidApproximation}
\begin{figure}
	\centering
	\includegraphics[width=\columnwidth]{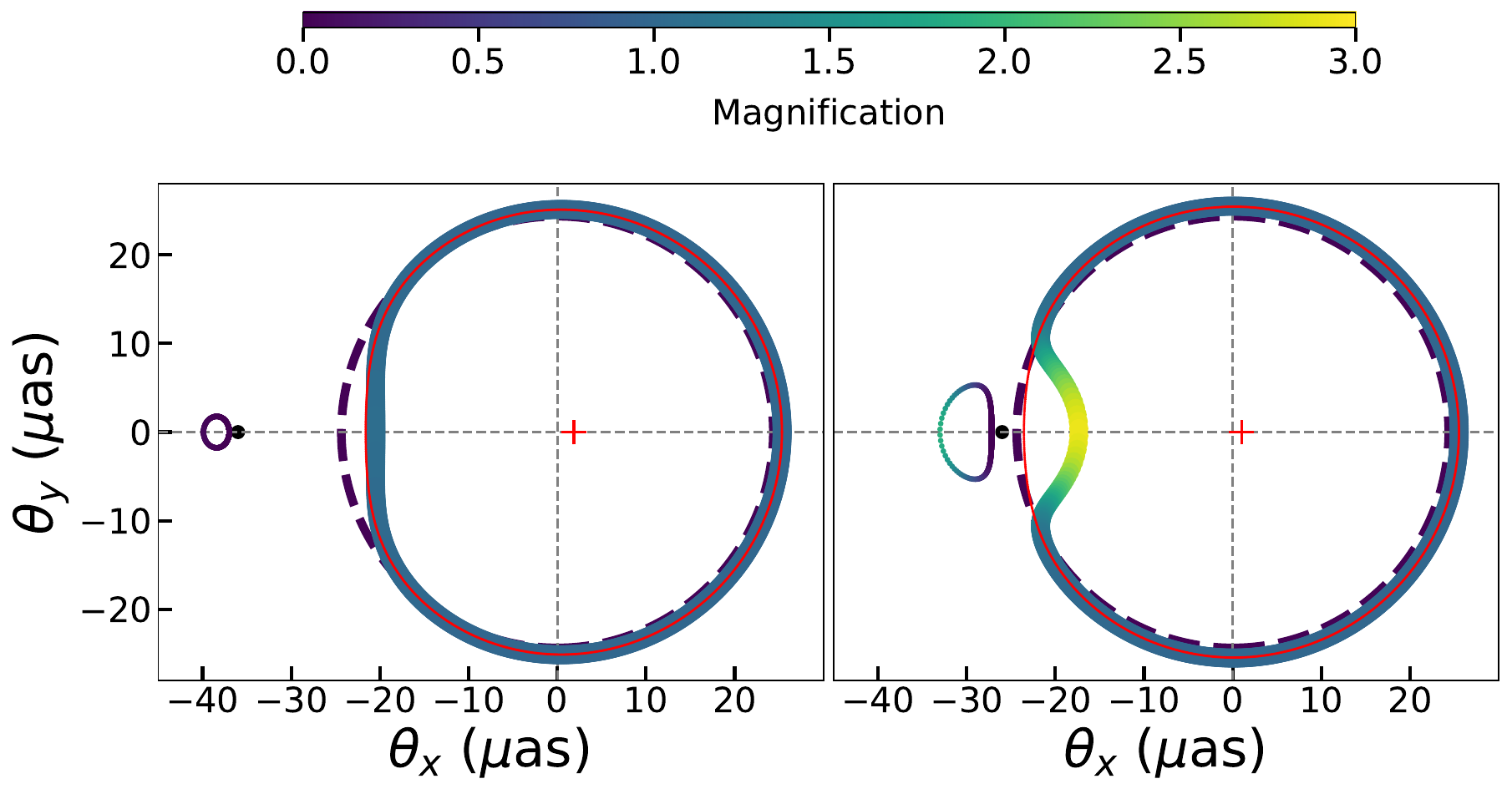}
	\caption{The figure displays two lensed images of the shadow of Sgr~A$^*$, with contours color-coded to represent the corresponding photometric magnification. The red contour represents the centroid approximation. The lens and source parameters are consistent with those in fig.\ref{fig:lensedshadow}. In the left panel, the angular separation of the lens is large, resulting in significant de-magnification of the secondary lensed image, leading to the primary lensed image predominantly contributing to the measured flux. The primary lensed image closely matches the centroid-approximated contour in this scenario. In the right panel, the lens is close enough for a portion of the shadow boundary to lie within $0.5\theta_E$ around the lens. Under such conditions, the secondary image becomes observable if the angular resolution is finer than $\theta_E$.}
	\label{fig:CentroidApproximation}
\end{figure}
It should be noted that fig.~\ref{fig:lensedshadow} shows the microlensed shadow under the centroid approximation. In principle, each point source on the boundary of the true shadow will have two lensed images due to the point lens as discussed in sec.~\ref{subsec:MLreview}. Hence, there will exist two lensed images of the shadow at any instant. In fig.~\ref{fig:CentroidApproximation}, we illustrate two lensed images of the shadow, color-coded to represent their corresponding photometric magnifications, along with the microlensed shadow under the centroid approximation outlined by the red contour. The lens and source parameters are still same i.e. $D_s = 8.2$ kpc, $R = 24.35~\mu$as, $M = 1~M_\odot$, and $r = 0.5$~pc.

In the left panel of the figure, we consider a scenario where the lens (depicted as a black dot) is sufficiently distant from the true shadow (indicated by the blue dashed circle) such that the angular separation ($\beta$) between all points on the true boundary is larger than the Einstein angle ($\theta_E$), i.e., $\beta > \theta_E$. Under these conditions, the primary shadow to the right of the lens exhibits an approximately unit photometric magnification throughout its extent. Conversely, the secondary image, positioned to the left of the lens, experiences significant demagnification. This distinction arises from the fact that the photometric magnifications of the primary and secondary images, as given by eq.~\ref{eq:ImageMagnifications}, exhibit variations described by $\mu_+\sim 1+\theta_E^4/\beta^4$ and $\mu_-\sim\theta_E^4/\beta^4$ respectively when $\beta$ is much larger than $\theta_E$. When observing such a lensing scenario of the shadow with a telescope like EHT, the flux contribution from the secondary image becomes negligible. Consequently, the primary lensed shadow, to which the centroid shadow is approximated, emerges as the observable feature of interest.

On the other hand, it can be calculated from eq.~\ref{eq:ImageMagnifications}, when the lens is close enough to the shadow such that there is a portion of the true shadow boundary for which $\beta<0.5\theta_E$, viz. shown in the right panel, then the secondary image of that portion can be magnified and roughly would be separated from the primary shadow with $\sim \theta_E$. Hence, the secondary shadow can, in principle, be observable along with the primary shadow, except the telescope resolution per epoch is less than the Einstein angle.

In summary, the centroid approximation is valid as long as the Einstein angle is smaller than the resolution per epoch of the telescope. However, when the Einstein angle is larger than the resolution per epoch, the observability of an additional secondary shadow, along with the distorted primary shadow, necessitates the lens to be positioned sufficiently close to the true shadow such that $\beta < 0.5\theta_E$ for some point sources along the true shadow boundary to remain magnified (or at least not demagnified). Nevertheless, even with sub-Einstein-angle resolution, if $\beta >> \theta_E$, the highly demagnified secondary shadow can be disregarded, and the centroid image can be accurately approximated to the primary shadow. Therefore, to determine the maximum allowable $\beta$ for which the distortion is measurable, the centroid approximation suffices. Consequently, for the purpose of studying the detectability of this phenomenon, we will continue to employ the centroid approximation throughout the paper.

In the next section, we will characterize the astrometric microlensing signal imprinted on the boundary of the shadow.

\section{Characterization of the microlensing signal}
\label{sec:Char}

We first define the signals of microlensing of the shadow and then we study their dependence on the lens and the true shadow.

\subsection{The lensed center}
 Equation~\ref{eq:lcenter} provides the location of the lensed center $\vv{OC}$. If the angular separation of the lens from the true center is $\xi$ and the Einstein angle is $\theta_E$, then the expression of the location of the lensed center will be
\begin{eqnarray}
\label{eq:lcenterdetailed}
\vv{OC} &=& -\frac{\theta_E^2(2\theta_E^2 - R^2 + \xi^2)}{[2\theta_E^2 +(R - \xi)^2][2\theta_E^2 +(R + \xi)^2]}\vv{\xi}.
\end{eqnarray}
The numerator of the above expression suggests that in the case of the radius of the true shadow smaller than or equal to $\sqrt{2}$ times the Einstein angle (i.e., $R\leq\sqrt{2}\theta_E$), the shift in the center will be along the negative x-axis for all $\xi>0$. However, in the case of $R>\sqrt{2}\theta_E$, the shift in the center will be along the negative x-axis for the angular separation greater than equal to $\sqrt{R^2 - 2\theta_E^2}$; otherwise, it would be along the positive x-axis. The lensed-center will coincide with the true center for a perfect alignment ($\xi=0$) of the lens with the true shadow. This is because, for perfect alignment, each point source on the boundary of the shadow shifts symmetrically; hence, the average position of the point sources at the boundaries would be unaltered.

\begin{figure*}
	\centering
	\includegraphics[width=\textwidth]{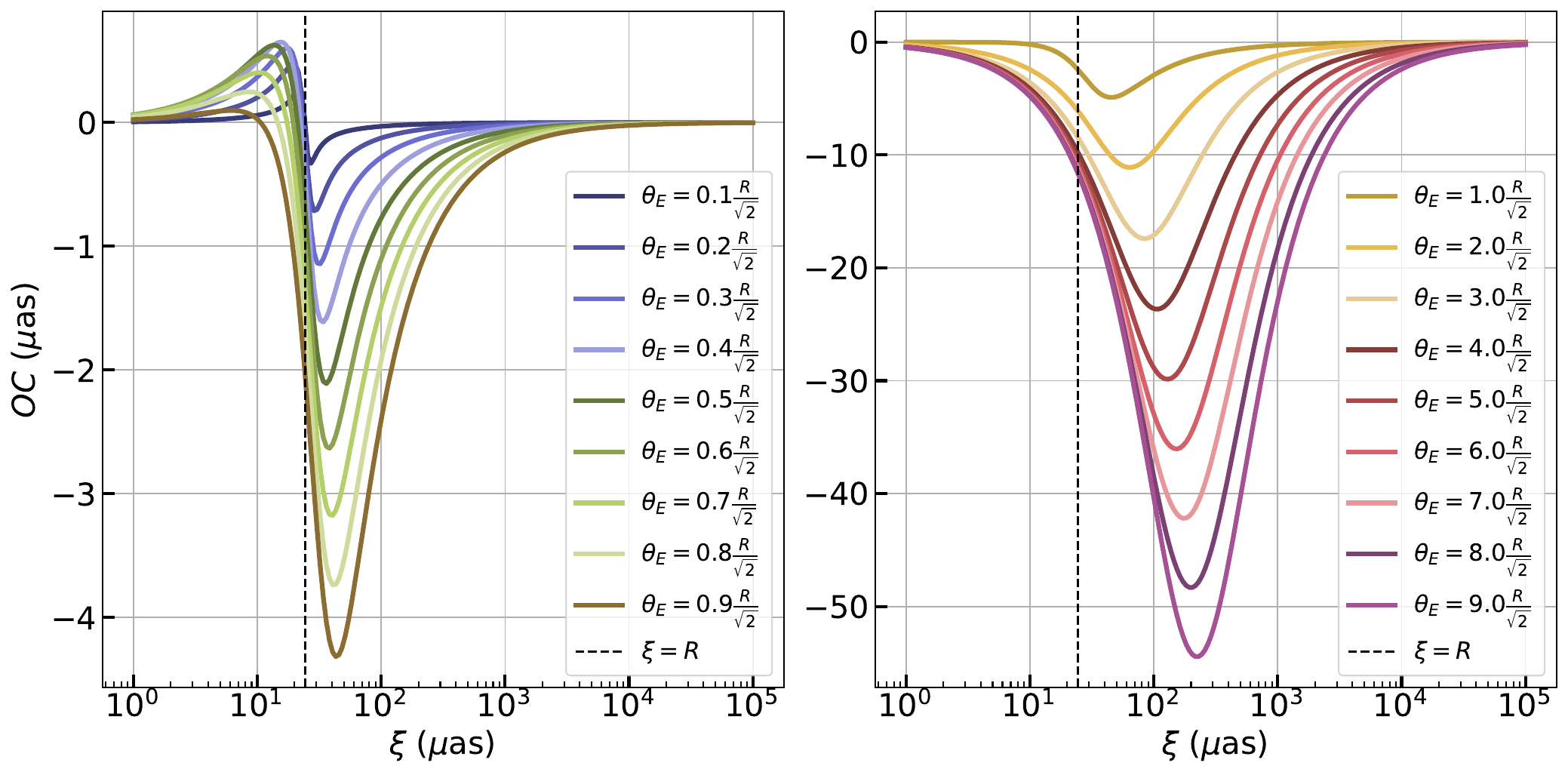}
	\caption{This figure shows how the angular position of the lensed-center $OC$, relative to the center of the true shadow, varies with the angular separation of the lens with the true center. The center is supposed to be moved in the direction of the lens when $OC$ has a positive value. However, a negative value of $OC$ shows the lensed-center shifts away from the lens. The radius of the true shadow, which we have taken to be of SgrA$^{*}$ ($R = 24.35~\mu$as) for demonstrative purposes, is shown by the vertical black-dashed line. In the \textit{left panel}, we plot $OC$ for the cases when the radius of the true shadow is greater than $\sqrt{2}\theta_E$. Every time the value of $\xi$ crosses the value of $\sqrt{R^2 - 2\theta_E^2}$, the sign of $OC$ changes. There are two extremum shifts in the center. One occurs when the lens is outside the shadow, and the associated extremum shift is away from the lens. The second occurs when the lens lies within the true shadow, and the corresponding extremum shift is towards the lens. For a small Einstein angle compared to the true size of the shadow, the shift in the center due to microlensing would be substantial only if the lens is close to the boundary of the true shadow. In the \textit{right panel}, we present the cases for which the radius of the true shadow is less than or equal to $\sqrt{2}\theta_E$. It is important to note that, in this case, the shift will always be away from the lens, and now, the extreme shift occurs only when the lens is outside of the true shadow.}
	\label{fig:lcenter}
\end{figure*}

In fig.~\ref{fig:lcenter}, we plot the magnitude of $\vv{OC}$, denoted by OC in the figure, with $\xi$. A negative value of $OC$ indicates that the lensed center has moved away from the lens due to microlensing. On the other hand, a positive value of $OC$ shows the lensed center is on the same side as the lens from the true center. A vertical black-dashed line marks the true boundary of the shadow, which is assumed to be of Sgr~A$^*$, at $R = 24.35~\mu$as. This figure suggests two important lessons about the location of the center of the apparent shadow. First, if the radius of the true shadow is smaller than $\sqrt{2}\theta_E$, the lensed center will always be on the opposite side of the lens from the true center. In contrast, there will be a sign change in $OC$ if the radius of the true shadow is greater than $\sqrt{2}\theta_E$. The sign change in $OC$ occurs near the boundary of the true shadow, which is evident from eq.~\ref{eq:lcenterdetailed}. Intuitively, this behavior becomes straightforward when we recall that the astrometric shift of any point source at the true boundary of the shadow is always away from the lens. This shift increases as $\beta$ increases, reaching its peak at $\beta = \sqrt{2}\theta_E$, and then gradually decreases with further increases in $\beta$. Consequently, a lens located outside the true shadow will consistently push all the point sources on the boundary away from the lens, causing an overall shift of the shadow away from the lens. This holds true regardless of the size of the shadow relative to the Einstein angle. However, for a lens within the true shadow, the shift can be towards the lens when the closer side of the true boundary shifts more than the far side, which is possible only when $R<\sqrt{2}\theta_E$. Hence, for lenses with a tiny Einstein angle as compared to $R$, the lensed center will be on the opposite side of the lens from the true center if the lens is outside of the true shadow, whereas it will be on the same side of the lens if the lens is within the true shadow. Second, there are angular positions of the lens for which there will be an extreme shift in the center. It can be seen from the figure there will be two $\xi$ values for which shift attains extremum values if Einstein angle is smaller than $R$ (specifically $R>\sqrt{2}\theta_E$). One extremum is attained when the angular position of the lens is outside of the true shadow, and the other occurs when the lens's angular position is inside the true shadow.

We also draw the same curve for various Einstein angle values relative to the shadow's true radius. The vertical black-dotted line represents the true boundary of the shadow at $\xi = R$. In the case of $\theta_E\geq R/\sqrt{2}$, typically, the shift grows with increasing angular separation. However, it attains a maximum value when the angular separation is around  $\sqrt{2}\theta_E$, and later as the angular separation is larger than this value, the shift goes down. The shift scales as $\theta_E^2/\xi^2$ for large angular separation ($\xi>>\theta_E$). In the case of $\theta_E<R/\sqrt{2}$, the shift has a non-trivial dependence near the boundary of the shadow. It changes sign near $\xi = \sqrt{R^2 - 2\theta_E^2}$, evident from the expression given by eq.\ref{eq:lcenterdetailed} as well. 

It should be noted that the absolute shift in the center of the shadow is not an observable signal in a single epoch of the image. Because if the shadow is microlensed, then the true center would not be observable. Therefore, a single frame of the shadow captured at any epoch would be unable to measure the shift in the center due to microlensing. However, one can observe the change in the shifted center if at least two frames of the shadow are taken at two different epochs. Therefore, the change in the position of the lensed-center between the two frames could, in principle, be an observable lensing signal. In practice, accurate astrometry on the shadow center is crucial to measure the shift in the center, which necessitates the use of a standard source in the field of view to act as a stable reference point. However, finding a well-calibrated source with precisely known astrometry towards the galactic center is extremely difficult due to highly crowded stellar density and high extinction, for eg.~\citep{2021A&A...649A...2L}. Furthermore, in the case of the duration of the microlensing event being much longer than the observation time, even any change in the lensed center would be unobservable.

\subsection{Variation of shadow radius with position angle}
\label{subsec:Signal}
\begin{figure*}
	\centering
	\includegraphics[width=\textwidth]{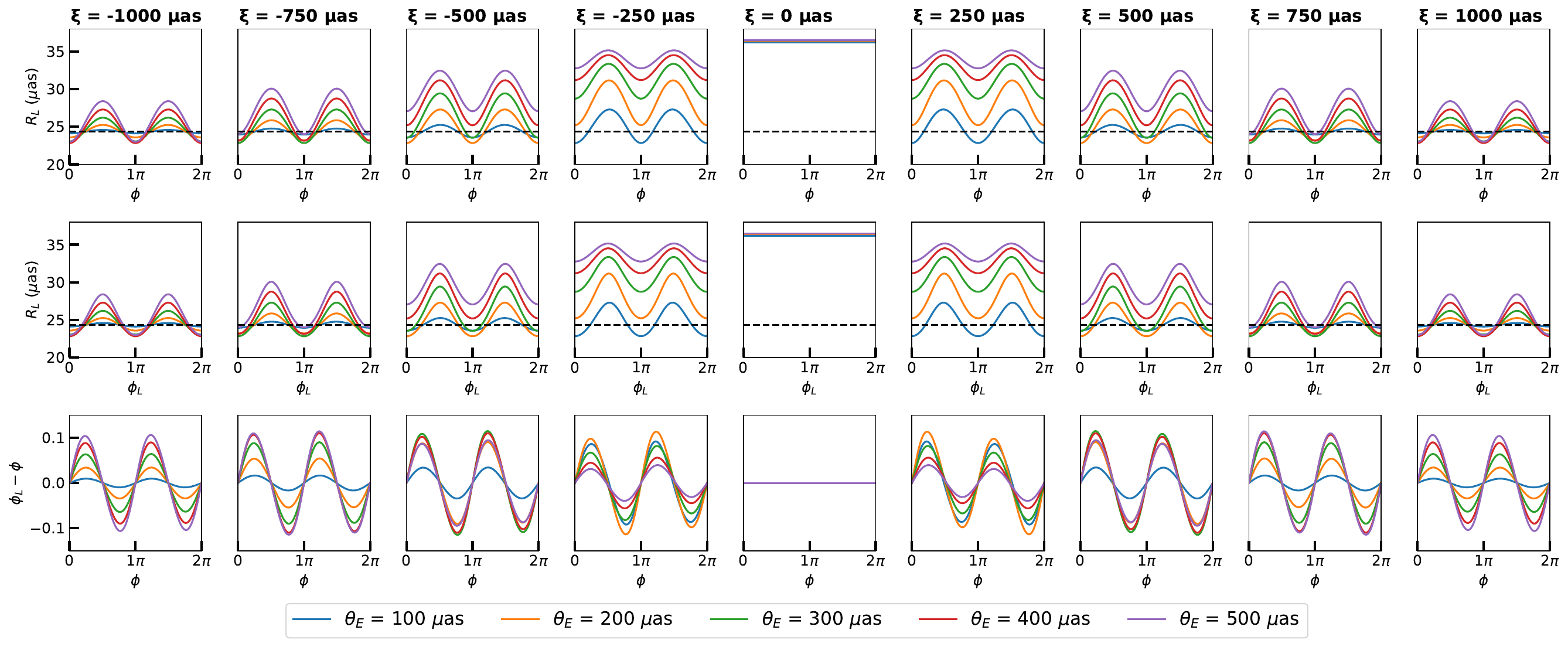}
	\caption{We describe the profile of the angular radius of the microlensed shadow with position angle. Here we show the angular radius profile when the Einstein angle is larger than the angular radius of the shadow. The horizontal dashed-line shows the true angular radius of Sgr~A$^*$. Each column corresponds to different angular separations of the lens with the center of the true shadow. The angular separation reduces from left to right, coincides with the center of the shadow at the central panel, and increases again. In the \textit{first row}, we plot the observed angular radius with the position angle of the true shadow. The \textit{second row} shows the observed profile of the angular radius of the microlensed shadow with the observed position angle. However, the \textit{third row} is drawn to show the difference between the apparent position angle and the true position angle of a point of the shadow and how it changes with the true position angle. The typical profile is well explained by eq.~\ref{eq:RLlargethE} and eq.~\ref{eq:phiLlargethE}.}
	\label{fig:RLphiLwithxi}
\end{figure*}
\begin{figure*}
	\centering
	\includegraphics[width=\textwidth]{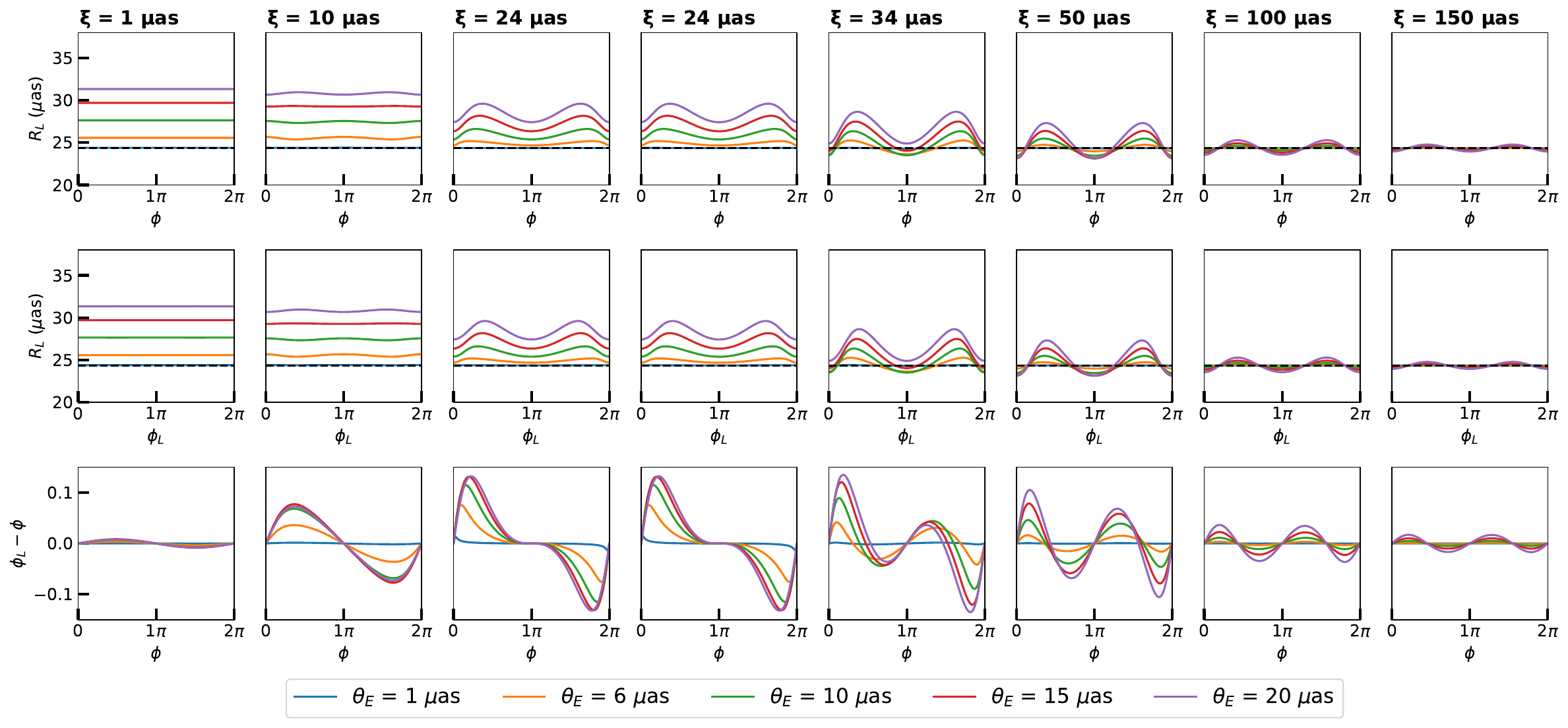}
	\caption{This figure shows the variation of the angular radius profile of the microlensed shadow similar to fig.~\ref{fig:RLphiLwithxi} but when the Einstein angle is smaller than the angular radius of the true shadow. Here we increase the angular separation across the columns from left to right. The basic feature of the profile can be encapsulated in eq.~\ref{eq:RLsmallthE} and eq.~\ref{eq:phiLsmallthE} (for more details see the text).}
	\label{fig:Signalnear}
\end{figure*}

Another feature of microlensing of the shadow is the variation of the radius with the position angle. A lensing signal is defined as the variation of the lensed radius $R_L$ with respect to the observed position angle $\phi_L$ at any instant. The exact dependence of $R_L (\phi)$ and $\phi_L (\phi)$, given by eq.~\ref{eq:RL} and eq.~\ref{eq:phiL} respectively, on the Einstein angle and the angular separation is not trivial to get an overall physical understanding. Therefore, we break our explanation into two cases (1) a much larger Einstein angle compared to $R$  and (2) a much smaller Einstein angle compared to $R$.

Firstly, we take the case when Einstein angle is much larger than the true angular radius of the shadow, i.e., $\theta_E>>R$. The radial profile of microlensed shadow can then be approximated as
\begin{eqnarray}
\label{eq:RLlargethE}
R_L &\approx&R_0 - A_0\cos{2\phi}
\end{eqnarray}
where the mean radius $R_0$ and the amplitude $A_0$ of the oscillation are given by
\begin{eqnarray}
R_0 &=& 
\begin{cases} 
\label{eq:R0largethE}
\frac{3}{2}R - \frac{R}{4}\frac{R^2}{\theta_E^2} - \frac{R}{2}\frac{\xi^2}{\theta_E^2}&; \xi<\theta_E, \\
R &;  \xi>\theta_E; 
\end{cases}\\
A_0 &=& 
\begin{cases} 
\label{eq:A0largethE}
 \frac{R}{4}\frac{\xi^2}{\theta_E^2}&; \xi<\theta_E, \\
 R\frac{\theta_E^2}{\xi^2}&; \xi > \theta_E; 
\end{cases}
\end{eqnarray}
and the observed position angle is approximated by 
\begin{eqnarray}
\label{eq:phiLlargethE}
\phi_L &\approx& 
\begin{cases} 
\phi + \frac{\xi R}{3\theta_E^2}\sin(\phi) +\frac{\xi^2}{6\theta_E^2}\sin(2 \phi)&; \xi<\theta_E, \\
\phi + \frac{\theta_E^2}{\xi^2}\sin(2\phi)&; \xi > \theta_E. 
\end{cases}
\end{eqnarray}
In fig.~\ref{fig:RLphiLwithxi}, we show how the exact variation of the angular radius with the position angle changes as the lens-shadow separation $\xi$ changes. The black-dashed horizontal line represents the true radius $R$. The oscillation in the radius about a mean radius, as the position angle changes from 0 to $2\pi$, is the primary characteristic of the subplots of the figure. We also show the dependence of the signal on the Einstein angle in the same figure. For a fixed $\xi$, it can be seen that the curve of $R_L$ with $\phi$ with a larger value of $\theta_E$ is always above the curve with a smaller value of  $\theta_E$. It implies that the average size of the shadow would be larger for the larger value of the Einstein angle for a given lens-shadow separation. The above approximate expression of $R_0$ captures this dependence explicitly. Furthermore, it is also evident from the dependence of $R_0$ that the average radius saturates to a value of $3R/2 - R^3/(4\theta_E^2)$ when the lens coincides with the true shadow shown in the subplot. Hence, microlensing cannot arbitrarily increase the size of the shadow with arbitrarily large $\theta_E$. In fact, if $\theta_E>>R$, then the maximum size attained by the microlensed shadow is $3R/2$ which is independent of $\theta_E$. Therefore, microlensing can only magnify the shadow up to 50\% of the true size of the shadow.

In contrast with the size of the shadow, there are oscillations in the radial profile, which are a signature of apparent asymmetry in the microlensed shadow. The oscillation amplitude increases as the lens moves away from the shadow, starting from $\xi=0~\mu$as (e.g., see the amplitude of the red curves in the subplots). It is also worth noticing that in the far region around the shadow, the oscillation amplitude also goes down as the lens asymptotically moves away from the lens. This asymptotic behavior is well captured by the $A_0$ expression given by eq.~\ref{eq:A0largethE}. It suggests that there will be an angular separation of the lens from the shadow for which the amplitude of the oscillation in the radius will be maximum and hence the asymmetry will be maximum. We numerically calculate this distance to be $\sim \sqrt{2}\theta_E$ in the limit of large Einstein angle\footnote{Notice that the shift of a centroid of a point source is maximum when the angular separation is $\sqrt{2}\theta_E$}. The maximum amplitude $A_{0,\textrm{max}}$ turns out to be $\sim R/8$ when $\xi \sim \sqrt{2}\theta_E$.

Secondly, we consider the case when the Einstein angle is much smaller than the true radius of the shadow, i.e., $\theta_E<<R$. The microlensed shadow profile can again be approximated by 
\begin{eqnarray}
\label{eq:RLsmallthE}
R_L &\approx&r_0 - a_0\cos{2\phi}
\end{eqnarray}
where the mean radius $r_0$ and the amplitude $a_0$ of the oscillations are given by
\begin{eqnarray}
r_0 &=& 
\begin{cases} 
\label{eq:r0smallthE}
R+\frac{\theta_E^2}{R}-2\frac{\theta_E^4}{R^3}&; \xi<R, \\
R&;  \xi>R; 
\end{cases}\\
a_0 &=&
\begin{cases} 
\label{eq:a0smallthE}
-\frac{\theta_E^2 \xi^2}{R^3}&; \xi<R, \\
 R\frac{\theta_E^2}{\xi^2}&; R < \xi; 
\end{cases}
\end{eqnarray}
and the observed position angle is approximated to be
\begin{eqnarray}
\label{eq:phiLsmallthE}
\phi_L &\approx& 
\begin{cases} 
\phi + \frac{2 \xi \theta_E^2}{R^3}\sin(\phi) +\frac{\xi^2\theta_E^2}{R^4}\sin(2 \phi)&; \xi<R, \\
\phi + \frac{\theta_E^2}{\xi^2}\sin(2\phi)&; R < \xi.
\end{cases}
\end{eqnarray}
In fig.~\ref{fig:Signalnear}, we plot an exact dependence of the $R_L$ on $\phi_L$ for various values of $\xi$. Again the black-dashed horizontal line represents the true radius $R$. The mean radius $r_0$ is smaller for a smaller value of Einstein angle, and asymptotically it coincides with the true radius when $\theta_E \rightarrow 0$. However, the amplitude of the oscillation $a_0$ increases as the lens moves away from the center of the shadow. Furthermore, for very large $\xi$, it again decreases. These behaviors of $r_0$ and $a_0$ are well described by eq.~\ref{eq:r0smallthE} and eq.~\ref{eq:a0smallthE} respectively.
\subsection{Magnification in size}
\label{subsec:size}
We define the radius of the microlensed shadow as the average of $R_L(\phi)$ over $\phi$. Hence, the average radius is given by
\begin{eqnarray}
\langle R_L \rangle&=& \frac{1}{2\pi}\int_0^{2 \pi} |\vv{R}_L(\phi)|d\phi.
\end{eqnarray}
In the limiting cases of $\theta_E>>R$ and $\theta_E<<R$, we have obtained an approximate expression of $R_L$ in the previous sec.~\ref{subsec:Signal}. Hence, the average radius in various limiting cases can be summarized as 
\begin{eqnarray}
\label{eq:avgRLlimit}
\langle R_L \rangle &=& 
    \begin{cases}
    R_0&; \theta_E>>R, \\
    r_0&;  \theta_E<<R; 
    \end{cases}
\end{eqnarray}
where $R_0$ and $r_0$ are given by eq.~\ref{eq:R0largethE} and eq.~\ref{eq:r0smallthE} respectively.

\begin{figure}
	\centering
	\includegraphics[width=\linewidth]{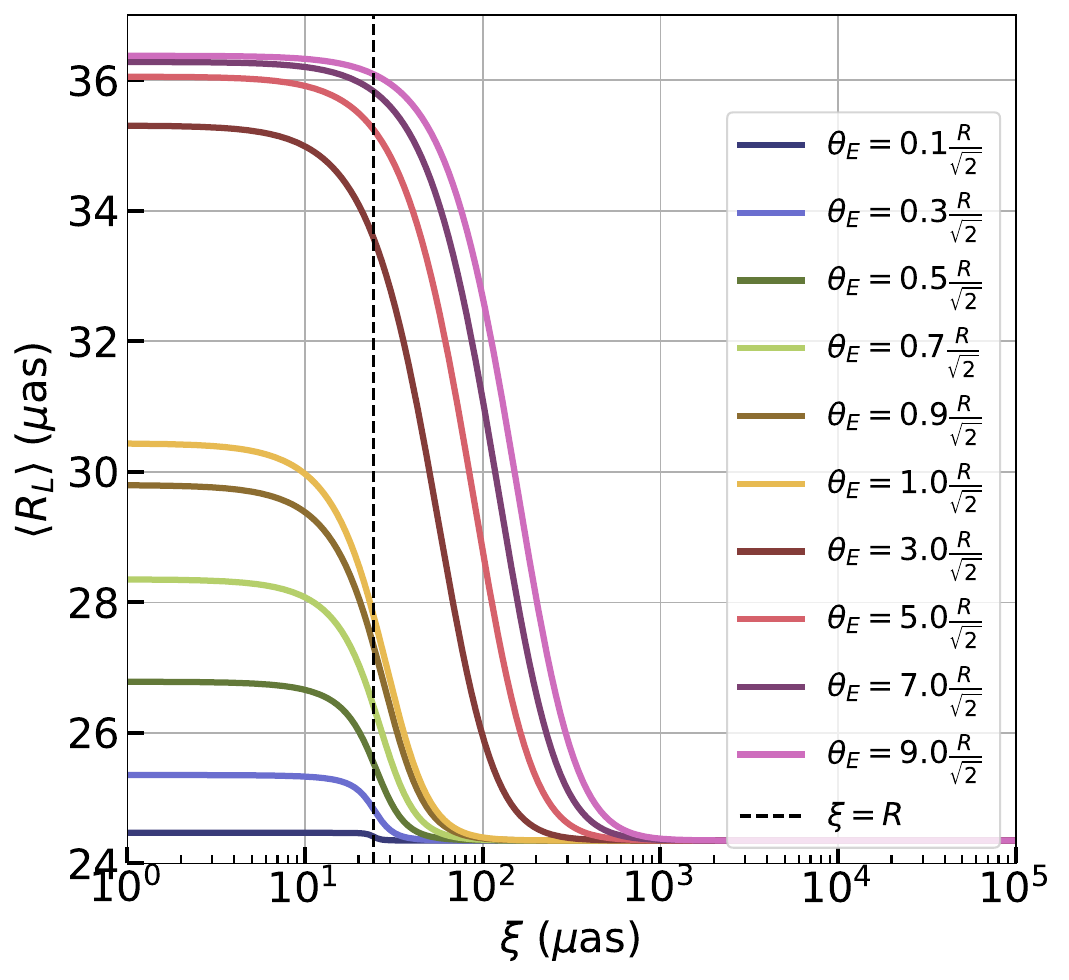}
	\caption{This figure depicts how the average angular radius changes as the angular position of the lens changes. The averaged angular radius approaches a maximum value as the angular separation of the lens is $\lesssim \theta_E$. It also shows that microlensing can not arbitrarily magnify the shadow. The maximum enlargement of the shadow that can occur through microlensing is 1.5 times the size of the true shadow when the Einstein angle is much larger than the true shadow.}
	\label{fig:size}
\end{figure}
In fig.~\ref{fig:size}, we show the change in the average size of the microlensed shadow with respect to the true size of the shadow as a function of the angular separation of the lens $\xi$. It can be seen from the plot that the size of the microlensed shadow gets magnified as the lens comes closer to the shadow, but there is a maximum magnification in size. There is a plateau region as the lens-shadow separation is less than around $\sqrt{2}\theta_E$. In the limit of $\theta_E>>R$ and $\xi<<\theta_E$, the leading order behavior of $R_L-R$ would be $R_0\sim R/2$. Hence, the maximum change in the size of the shadow will be half of the true size of the shadow. However, in the limit of $R>>\theta_E$ and $\xi<<R$, the plateau goes down as the Einstein angle reduces, which can also now be understood with the leading order behavior of the $R_L-R$ as $r_0 \sim \theta_E^2/R$. The dependence of the change in the size of the shadow on the explicit parameters of the lens such as mass and the distance of the lens from the center of the Milky Way can also be seen from fig.~\ref{fig:ObservableRadius} in the appendix.

\subsection{Asymmetry in shape}
\label{subsec:asymmetry}

\begin{figure*}
	\centering
	\includegraphics[width=\textwidth]{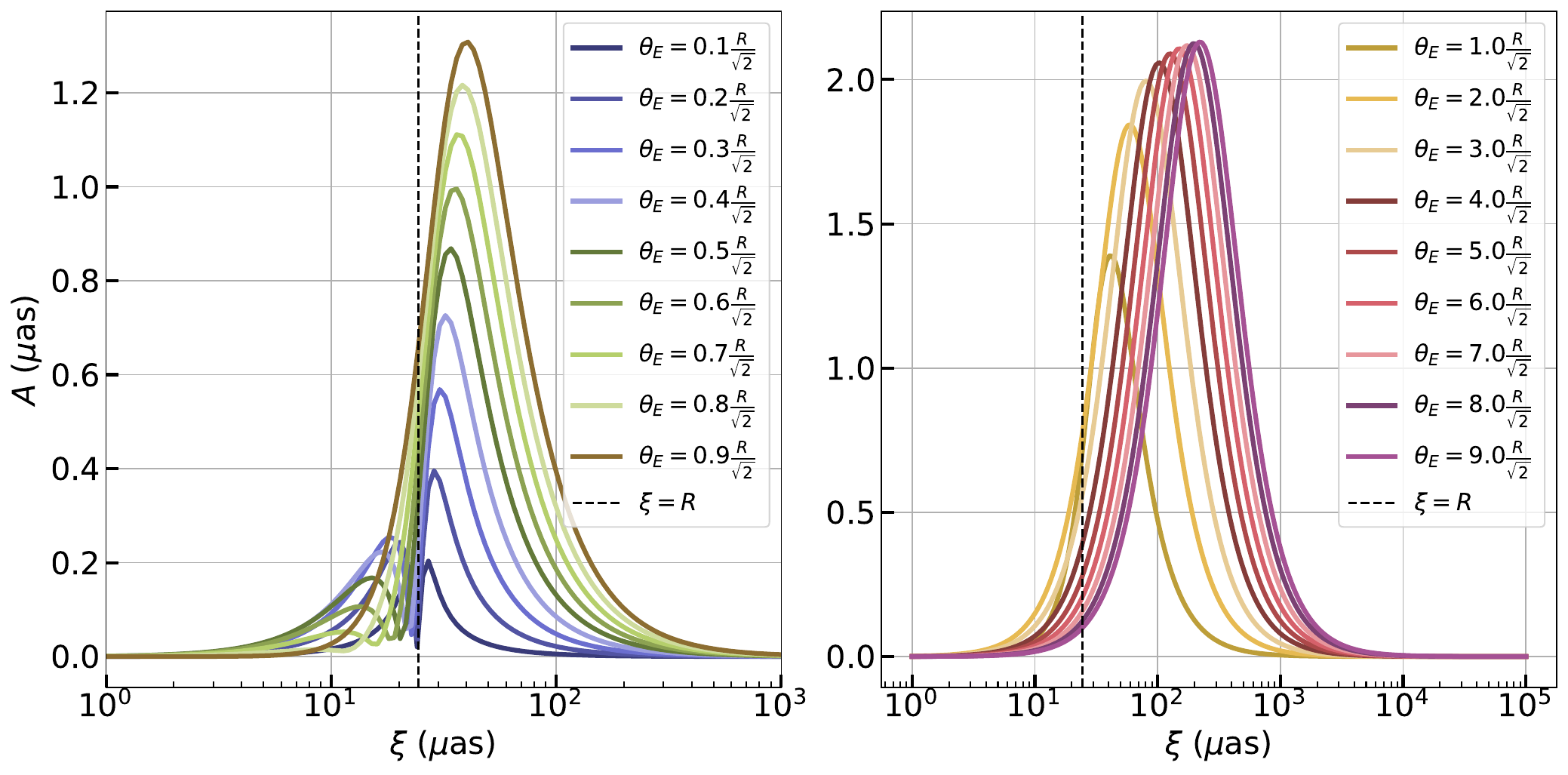}
	\caption{This figure depicts the variation of the asymmetry in the microlensed shadow with the lens angular separation, which is calculated numerically using the exact expressions given by eq.13 and eq.26. Different curves are drawn for different Einstein angles represented with different colors. We separate the asymmetry signal in the above two panels depending upon the relative size of the lens relative to the black hole shadow. The dashed-vertical line is the angular radius of the true shadow, which is again taken to be of Sgr~A$^*$. \textit{Left panel} shows the case of Einstein angle less than $R/\sqrt{2}$, which approximately implies a smaller Einstein angle as compared to the angular radius of the true shadow. There are two maxima in the asymmetry, the first maxima occurs when the lens is outside of the boundary of the true shadow, and the second occurs when the lens is within the shadow's boundary. For a sufficiently small Einstein angle, the asymmetry signal is mostly strong when the lens is near the true shadow's boundary. \textit{Right panel} is the asymmetry signal for a larger Einstein angle as compared to the size of the shadow ($\theta_E>R/\sqrt{2}$). In this case, the maximum asymmetry is obtained for the angular lens position outside the true shadow. Asymptotically for a very large Einstein angle, the maximum asymmetry approaches the order of the radius of the true shadow.}
	\label{fig:Asymmetry}
\end{figure*}

The oscillation in the radius versus position angle curve is a signature of the asymmetry in the shadow, as shown in fig.~\ref{fig:RLphiLwithxi} and fig.~\ref{fig:Signalnear}. Unlike the shift in the center and the magnification in the size of the shadow, asymmetry is an absolute observable effect of the microlensing of shadow. Hence, it can be measured even in a single epoch frame of the shadow. The asymmetry of the lensed shadow is defined as the variance in the observed radius of the shadow and given by,
\begin{eqnarray}
	A &=& \sqrt{\frac{1}{2\pi}\int_0^{2\pi} (R_L(\phi)^2 - \langle R_L \rangle^2)d\phi}.
\end{eqnarray}
The asymmetry in the various limits can be summarized by
\begin{eqnarray}
A &=& 
    \begin{cases}
    \frac{A_0}{\sqrt{2}}&; \theta_E>>R, \\
    \frac{a_0}{\sqrt{2}}&;  \theta_E<<R; 
    \end{cases}
\end{eqnarray}
where the oscillation amplitudes $A_0$ and $a_0$ are given by eq.~\ref{eq:A0largethE} and eq.~\ref{eq:a0smallthE} respectively.

In fig.~\ref{fig:Asymmetry}, we show the exact dependence of the asymmetry on the angular separation $\xi$ and the Einstein angle $\theta_E$. The curves with different colors signify the various values of the Einstein angle. There are two distinct behaviors of the curve depending upon whether the true radius of the shadow is less than $\sqrt{2}\theta_E$ or greater than $\sqrt{2}\theta_E$. 

In the case of $R<\sqrt{2}\theta_E$, shown in the right panel of  fig.~\ref{fig:Asymmetry}, the asymmetry increases with an increase in $\xi$, reaching a maximum value at $\xi \approx \sqrt{2} \theta_E$, and then goes down with a further increase in the angular separation $\xi$. This is because, for the large Einstein angle, the amplitude $A_0$ is given by eq.~\ref{eq:A0largethE} which for small $\xi$ varies as $\sim \xi^2/ \theta_E^2$. However, when the lens is far from the shadow ($\xi>\theta_E$ ), $A_0\sim \theta_E^2/\xi^2$ and  decreases as we increase the lens-shadow separation. The approximate separation for which there will be the maximum asymmetry is obtained numerically to be $\sim \sqrt{2}\theta_E$ and the corresponding maximum asymmetry ever achieved due to the microlensing effect is found to be
\begin{eqnarray}
A_\textrm{max}&\approx& \frac{R}{8\sqrt{2}},
\end{eqnarray}
which is independent of the Einstein angle itself. This explains why the maximum asymmetry in fig.~\ref{fig:Asymmetry} is the same for all $\theta_E>>R/\sqrt{2}$.

On the other hand, two maxima occur in the case of $R>\sqrt{2}\theta_E$ which is shown by the left panel of the figure. The first maxima occurs when the lens is within the true shadow, and another occurs outside. In this limit of a small Einstein angle compared to the true size of the shadow, the asymmetry signal would be significant when the lens is near the boundary of the shadow,  also shown in the left panel of the figure. 

\section{Detectability of the microlensed shadow in an EHT-like telescope}
\label{sec:detectability}
We shall now estimate the telescope specifications needed to detect the microlensing of a black hole shadow. This requires quantifying, first, the uncertainty in measuring the shape of the shadow with a given telescope specification and second, how often a lens will be aligned with the shadow to induce a detectable signal. To demonstrate the procedure, we shall again concentrate on the microlensing effect on Sgr~A$^*$ for which the abundance of stellar lenses in the Milky Way are well constrained observationally. 

Apart from  the shift in the center, the enlargement in the average size, and the asymmetry in the microlensed shadow, we saw in the previous sec.~\ref{sec:Char} that the asymmetry is an absolute observable (assuming the circular form of the true shadow). Additionally, the average size of the shadow of Sgr~A$^*$ is tightly constrained with precise measurements of its distance and mass by the stellar orbits around Sgr~A$^*$~\citep{2016ApJ...830...17B}; therefore, if the measured size surpasses the expected size from the stellar orbits, it indicates an observable effect of microlensing. The approximate expressions for the asymmetry and the average size up to the leading order in $\xi$ are derived under the assumption that the lens is far from the boundary of the shadow compared to the Einstein angle. They are given by
\begin{eqnarray}
    A  &\approx& \frac{R}{\sqrt{2}}\frac{\theta_E^2}{\xi^2}
\end{eqnarray}
and 
\begin{eqnarray}
    \langle R_L \rangle &\approx& R \left(1+ \frac{9}{4}\frac{\theta_E^4}{\xi^4}\right).
\end{eqnarray}
Considering the uncertainty in measuring the shadow radius as $\sigma_a$, which represents the standard deviation of the radial measurement, we determine the maximum threshold angular separation $\xi_\textrm{th}$ of the lens from the center of mass of Sgr~A$^*$ such that $A=\sigma_a$ and $\langle R_L \rangle - R = \sigma_a$, as follows:
\begin{eqnarray}
\xi_{\textrm{th,}A} &=& 
    26.6 \theta_E \left(\frac{10^{-3}}{\sigma_a/R}\right)^{1/2},
\end{eqnarray}
for asymmetry and 
\begin{eqnarray}
\xi_{\textrm{th,}R_L} &=& 
    0.7 \theta_E \left( \frac{10^{-3}}{\sigma_a/R}\right)^{1/4}
\end{eqnarray}
for size magnification. These thresholds define the angles within which a lens must lie to induce a detectable microlensing event corresponding to each observable. As the asymmetry observable dies out more slowly ($\theta_E^2/\xi^2$) compared to the size magnification ($\theta_E^4/\xi^4$) with increasing lens angular separation $\xi$, the asymmetry observable will give us a maximum event rate due to the larger threshold angle compared to the size magnification. There will be a subset of the maximum number of events, which will show a detectable magnification in size. We will see the event rate scales linearly with $\xi_\textrm{th}$, and the threshold separation for the two observables are related by $\xi_{\textrm{th,}A} \approx 38 \left( 10^{-3}/(\sigma_a/R) \right )^{1/4} \xi_{\textrm{th,}R_L}$. Hence, we utilize the asymmetry observable as a proxy to determine the total microlensing event rate. The subset of events inducing detectable size magnification can then be obtained by dividing the total event rate with the factor $38\left(10^{-3}/(\sigma_a/R)\right)^{1/4}$.

\begin{figure*}
	\centering
	\includegraphics[width=\textwidth]{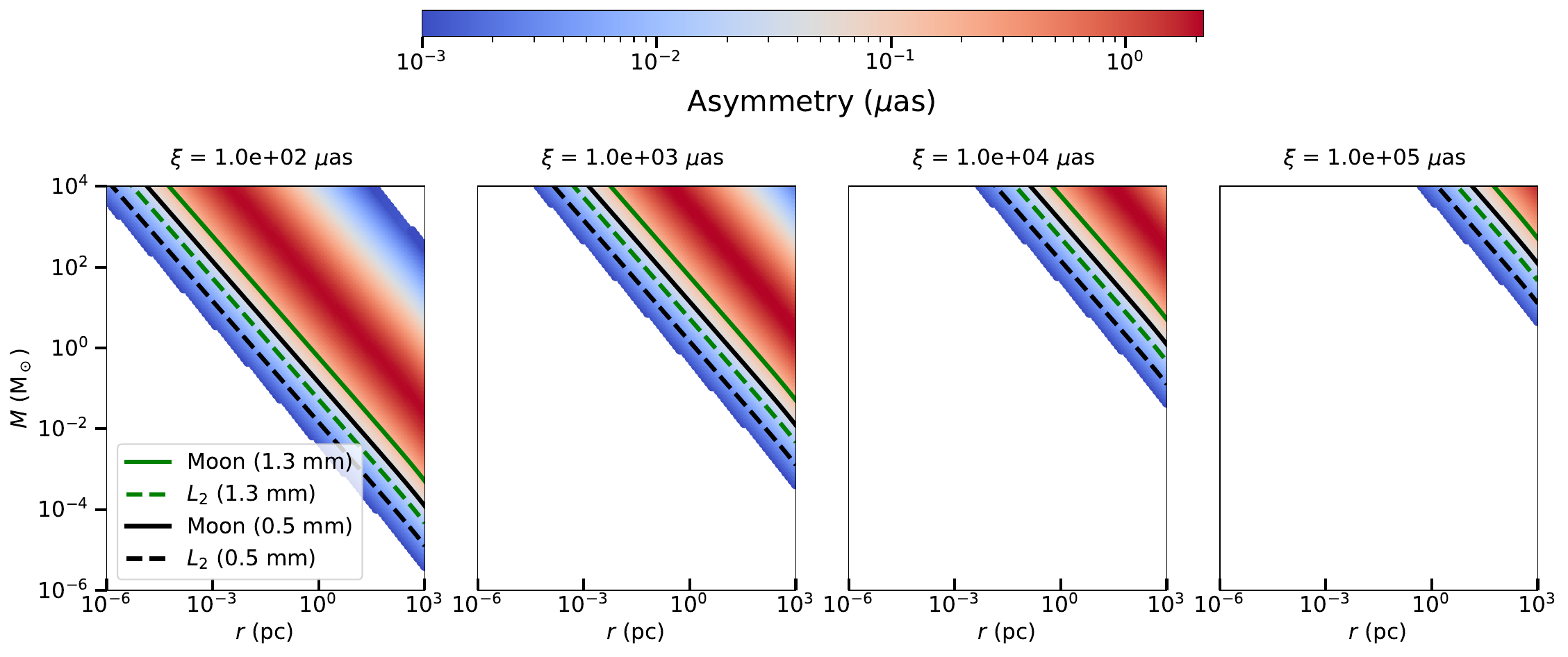}
	\caption{This figure shows the parameter space of the lens to be probed by various possible radio baselines in space by observing the shadow of Sgr~A$^*$. The color represents the asymmetry induced in Sgr~A$^*$ due to microlensing by a lens, in our line-of-sight, of a given mass and distance from the center of the Milky Way. Various panels show how the asymmetry and hence the region of the parameter space changes as the angular separation $\xi$ of the lens changes with distance from the shadow. The region above solid lines and dashed lines contours are the required characteristics of the lens for showing asymmetry signals in various baselines at Moon and $L_2$ positions, respectively. The black and green contours correspond to 1.3 mm and 0.5 mm operating radio wavelengths, respectively.}
	\label{fig:DetectableParams}
\end{figure*}

According to sec.~\ref{subsec:asymmetry}, the asymmetry in the shadow due to microlensing results in variance ($\sigma_a^2$) in the radial measurements of the shadow. Regardless of how large the Einstein angle is, the maximum standard deviation in the radius of Sgr~A$^*$ can be up to $A_\textrm{max}= 2.152~\mu$as due to microlensing. Therefore, the radius of the shadow cast by Sgr~A$^*$ must be measured with a precision of at least $2.152~\mu$as in order to have any microlensing signal. The current Event Horizon Telescope has measured the diameter of the shadow to be $2 R = 48.7\pm7.0~\mu$as~\citep{2022ApJ...930L..12E}. Hence, there 
cannot be any microlensing signature in the currently available image of Sgr~A$^*$ by EHT. However, given the future potential of EHT and the planned space-based EHT-like VLBI facilities~\citep{Roelofs:2019nmh, 2020AdSpR..65..821F, Mikheeva:2020bqj, Gurvits:2022wgm, Tiede:2022grp, Chael:2022meh, 2022AcAau.196...29R}, we anticipate achieving direct imaging of Sgr~A$^*$ with a sub-microarcsec level of resolution. As a result, it should be  possible to observe the microlensing phenomena that we suggest via direct imaging of the SMBH shadow.

We can estimate the projected precision in the radial measurements that could be made at EHT-like VLBI facilities in the near future. These estimates will allow us to find the range of mass and distance of the lens, which may result in an observable microlensing effect on the black hole shadow of Sgr~A$^*$. 

Let us assume that the largest baseline that can be used is $D$ and that the radio observation's wavelength is $\lambda$. This will enable the maximum resolution of the black hole shadow to be $\theta_\textrm{res}\sim\lambda/D$ per epoch. We consider an annular region with an angular radius of $R$ and an angular thickness of $\theta_\textrm{res}$. Hence, the number of data points $N$ on the boundary of the shadow is given by,
\begin{eqnarray}
N &\equiv& \frac{2\pi R\theta_\textrm{res}}{\pi\theta_\textrm{res}^2} = \frac{2R}{\theta_\textrm{res}}.
\end{eqnarray}
This implies that we will have N number of radial measurements of the shadow corresponding to the $N$ directions $\phi$ in the interval of $\phi\in[0,2\pi)$ with step size $\frac{2\pi}{N}$. Each radial measurement will have an accuracy of $\theta_\textrm{res}$. Therefore, assuming the true boundary of the shadow to be circular, the standard deviation of the angular radii about the true radius $R$ would be
\begin{eqnarray} 
\sigma_a &=& \frac{\theta_\textrm{res}}{\sqrt{N}} \nonumber \\
&\approx& 0.0134~\mu\textrm{as} \sqrt{\left(\frac{\lambda}{1\textrm{ mm}} \frac{10^6 \textrm{ km}}{D}\right)^3 \frac{24.35 \mu\textrm{as}}{R}}.
\end{eqnarray}
 The percent accuracy achieved will then be given by
\begin{eqnarray}
\%\textrm{error} &\equiv& \frac{\sigma_a}{R}\times 100 \nonumber\\
&\approx& 0.055 \% \left(\frac{\lambda}{1\textrm{ mm}} \frac{10^6 \textrm{ km}}{D} \frac{24.35 \mu\textrm{as}}{R}\right)^{3/2} .
\end{eqnarray}

We examine the detectability of the microlensing phenomenon in the black hole shadow of Sgr~A$^*$ through three benchmark baseline configurations. First is \textit{Earth} configuration, where the maximum baseline length corresponds to the size of Earth, akin to the configuration employed by the EHT. Second, \textit{Earth-Moon} configuration considers a setup where a radio telescope is positioned on Earth, while another is situated on the Moon, creating a maximum baseline equivalent to the Earth-Moon distance. The third configuration \textit{Earth-L$_2$} consists of a radio telescope located at L$_2$ (the second Lagrange point of the Earth-Sun system) in conjunction with one on Earth, forming a maximum baseline separated by Earth-L$_2$. In each configuration, we consider two operational radio wavelengths, namely 1.3 and 0.5 mm. For the Earth-Moon and the Earth-L$_2$, we present the uncertainty in the radial measurements in each epoch because of the unknown cadence and \textit{imaging time} ($t_\textrm{im}$), where the imaging time is the required time to make one black hole image, which would depend upon how these arrays are implemented for these space-based baselines. However, for the Earth-based arrays currently available in EHT operating at 1.3 mm, we know the observation scheme. According to the EHT collaboration \citep{2022ApJ...930L..13E},  there are two 8-minute scans of Sgr~A* per hour, followed by $\sim$3-minute scans, with an 8-minute gap. This observation scheme has been followed for about a day to obtain one image, i.e. the imaging time for EHT is currently about a day. We present the accuracy achieved by these configurations of the possible baselines in tab.~\ref{tab:Uncertainty}.
\begin{table}
  \centering
  \begin{tabular}{@{}p{2.6cm}|p{0.7cm} p{0.7cm}p{0.7cm} p{0.7cm} p{0.7cm}@{}}
    \hline
	\hline
    \multicolumn{6}{c}{$\lambda = 1.3$ mm}\\
	\hline
	   $D$ (km)& $\theta_\textrm{res}$ & N & \# of& $\sigma_a$ & $\%$\\
	&  ($\mu$as)&  & epochs & ($\mu$as) & error\\
	\hline
	10,700 (Earth)  & 25.06  & 1.9 & 1 & 17.98 & 73.8 \\
	300,000 (Earth-Moon)& 0.89  & 54.5 & 1 & 0.121 & 0.50\\
	1,500,000 (Earth-L$_2$)& 0.18  & 272.4 & 1 & 0.011 & 0.04\\
	\hline
	\multicolumn{6}{c}{$\lambda = 0.5$ mm} \\
	\hline
	10,700 (Earth)& 9.64  & 5.1 &1 & 4.288 & 17.61\\
	300,000 (Earth-Moon)& 0.34  & 141.7 & 1 & 0.029 & 0.12\\
	1,500,000 (Earth-L$_2$)& 0.07  & 708.3 & 1& 0.003 & 0.01\\
	\hline
  \end{tabular}
  \caption{This table shows the uncertainty in the measurement of the radial measurements at various baseline configurations. It is used to make predictions for the observability of the asymmetry in the microlensed shadow of Sgr~A$^*$ due to the stellar granularity of the Milky Way.}
  \label{tab:Uncertainty}
\end{table}

Using these benchmark resolutions of various space-based baselines, we derive the in-principle range of mass and distance of the lens, which can potentially cause a detectable microlensing signature in direct imaging of Sgr~A$^*$. The values of mass and distance, which can potentially be probed with these possible baselines, are shown in fig.~\ref{fig:DetectableParams}. The exact detectability of the microlensed shadow would require a detailed analysis of the visibility amplitude as a function of baselines which could distinguish the microlensed shadow and the true shadow with a given number of baselines, flux sensitivity, exposure time, and the dirty beam modeling of a VLBI facility. This will require a full numerical signal analysis for a given telescope specification which we leave for future exploration. This work will assume a simplified detection criterion of $A>\sigma_a$.

\subsection{Event duration}
\label{subsec:EventDuration}
Generically, we expect a relative lens motion relative to the true shadow. Therefore, the microlensed shadow changes its center, size, and shape with time. In the case of a moving lens, we then define the \textit{event duration} as the duration for which the asymmetry is above the detection threshold $\sigma_a$. For the relevant parameter space of lenses as depicted by fig.~\ref{fig:DetectableParams}, we can estimate a sensitive region around Sgr~A$^*$ within which a lens has to lie to create a measurable asymmetry by various possible baselines. We can then estimate the maximum threshold value of $\xi$ by taking asymmetry in the large $\xi$ limit, as
\begin{eqnarray}
\xi_\textrm{th} &\approx& \theta_E\sqrt{\frac{R}{\sigma_a \sqrt{2}}}, \nonumber\\
&\approx& 26.5 \text{ mas }\sqrt{ \frac{M}{1 \textrm{ M}_{\odot}}\frac{8.2 \textrm{ kpc}}{D_s}\frac{1-x}{x}\frac{10^{-3}}{\sigma_a/R}}.
\end{eqnarray}

\begin{figure}
	\centering
	\includegraphics[width=\linewidth]{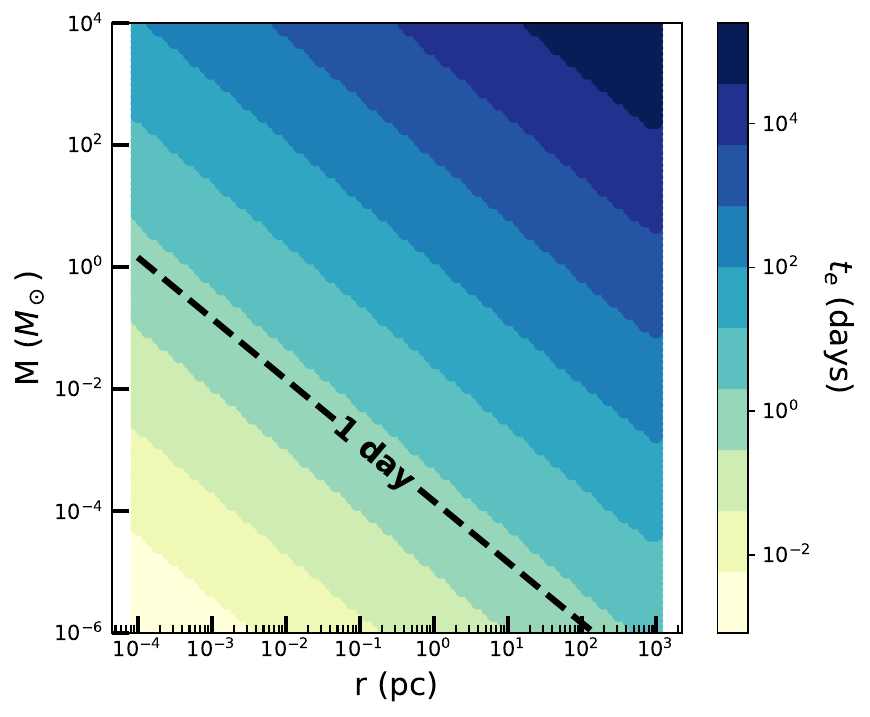}
	\caption{The above plot shows the duration of the microlensing event $t_e$ during which asymmetry remains above 0.1\% of $R$ and its dependence on the lens mass $M$ and the distance $r$ from Sgr~A$^*$ towards earth. The relative velocity of the lens is taken to be 100 km/sec with respect to the black hole shadow.}
	\label{fig:EventDurationanalysis}
\end{figure}

 Fig.~\ref{fig:EventDurationanalysis} shows the variation in the duration of the microlensing event with the characteristics of the lens. The relative velocity of the lens is taken to be 100 km/sec with respect to the black hole shadow.

Assuming the lens proper motion component along the radial direction of the true shadow is $\mu_r$, the event duration of the microlensing event is given by,
\begin{align}
t_e \equiv& \frac{2\xi_\textrm{th}}{\mu_r},\nonumber \\
\approx&  20.6 \text{ yr }\frac{100~\textrm{km/s}}{v}\left(\frac{D_s}{8.2 \textrm{ kpc}}\right)^{1/2}\sqrt{ \frac{M}{1 \textrm{ M}_{\odot}}x(1-x)\frac{10^{-3}}{\sigma_a/R}}.
\end{align}

It should be noted that to achieve the required contrast of interference fringes in radio measurements, one needs to integrate the interference visibility amplitude over some minimum exposure time (depending upon the brightness of the source),  and then repeated measurements over some time result in one black hole image. This collectively adds up to an effective imaging time $t_\textrm{im}$, and hence, an observed black hole image will always be time-averaged over $t_\textrm{im}$. Therefore, the duration for which the image of the shadow changes due to lensing should be larger than the imaging time. Otherwise, the microlensing feature would be washed out during the averaging. The imaging time used for Sgr~A$^*$ in the current EHT observation is $\sim1$~day~\citep{2022ApJ...930L..12E}. For the potential space-based configurations, it has been estimated the imaging time of a month can achieve the microarcsec-level resolution (see ref.~\citep{Gurvits:2022wgm}), and in that case the microlensing events of durations above a month will be only observable. The exact imaging time will depend on many factors that go into implementing the space-based EHT-like facilities, which is beyond the scope of this paper to discuss in detail\footnote{See ref.~\citep{2019A&A...625A.124R, 2020AdSpR..65..868G,2021ExA....51..559G, 2022AcAau.196...29R, 2022MNRAS.511..668L} for more details on the possible implementation of space interferometry.}. However, there is no fundamental limitation to reducing this imaging time, and it can be achieved with technological development. Hence, we will take a benchmark imaging time $t_\textrm{im}$= 1 day. We will thus assume that the target range of parameters of the lens, inducing lensing event with duration above one day (see e.g. dashed-line in fig.~\ref{fig:EventDurationanalysis} drawn for the accuracy in radial measurement to be 0.1\% of R), will cause microlensing signals which will not be washed out after averaging  over the imaging time of a day. Furthermore, we can even expect a time-varying change in the shape of the shadow during the passes of a lens in front of Sgr~A$^*$ with sufficient cadence and observational period of EHT-like facilities.

\subsection{Occurance of the microlensing due to stellar objects}
\label{subsec:EventRate}
The shadow of Sgr~A$^*$ is surrounded by various granular stellar objects. We estimate the probability that a microlensing event will be observed with a given baseline at any instant. This probability can be estimated to be $1-e^{-\tau} \approx \tau$, where $\tau$ is the optical depth given by~\citep{1994ApJ...430L.101K, Paczynski:1996nh},
\begin{eqnarray}    
\tau &=& \int_{0}^{D_s}\pi \xi_\textrm{th}^2\frac{\rho(D_l)}{M}D_l^2dD_l.
\end{eqnarray}

The stellar density $\rho(D_l)$ towards the Galactic center of the Milky Way can be split into stellar bulge $\rho_b(r)$ and stellar disc $\rho_d(r)$. The stellar bulge can be modeled with the Hernquist~\citep{1990ApJ...356..359H} density profile given by
\begin{eqnarray}
    \rho_b(r) &=& \frac{M_b}{2\pi}\frac{a}{r(r+a)^3}, 
\end{eqnarray}
where $r$ is the radial distance from the Galactic center, the mass parameter $M_b = 2\times 10^{10}~M_\odot$~\citep{2016A&A...587L...6V} and the scale radius $a$ = 0.31~kpc~\citep{2016arXiv161207781L}. The stellar disc is described by 
\begin{eqnarray}
    \rho_\textrm{disc}(r) &=& \frac{M_d}{8\pi z_0 r_d^2} e^{-r/r_d},
\end{eqnarray}
where $M_d = 4.8\times10^{10}~M_\odot$ is the disc mass, and $r_d = 2.67$~kpc is the disc scale length, and $z_0 = 0.32$~kpc is the scale height. Hence, the optical depth is found to be
\begin{eqnarray}
    \tau &\approx& 0.003 \frac{10^{-3}}{\sigma_a/R} \left(\frac{D_s}{8.2 \textrm{ kpc}}\right)^2.
\end{eqnarray}

We can also estimate the differential event rate of microlensing events using $d\Gamma \equiv d\tau/dt_e$. In section~\ref{subsec:EventDuration}, we established the significance of the duration $t_e$ in a microlensing event, emphasizing its necessity to exceed $t_\textrm{im}$. This ensures that the distinctive microlensing distortion signature remains discernible during the construction of a single image using data collected within the time frame $t_\textrm{im}$. Consequently, the total count of observable events within a specific baseline configuration relies on the distribution of events across different $t_e$ values. To quantify this, we calculate the annual event count within twenty logarithmically spaced $t_e$ bins spanning $ \log_{10}(t_i/\textrm{1 yr}) \in [-2,6]$. The events in the $i^\textrm{th}$ bins are counted using
\begin{eqnarray}
    \Delta \Gamma_i & =& \int_{t_i}^{t_f}\frac{\pi}{2} \xi_\textrm{th}\frac{v}{D_l}\frac{\rho(D_l)}{M}D_l^2\frac{dD_l}{dt_e} dt_e,
\end{eqnarray}
where $t_f = t_i\times 10^{0.4}$.

\begin{figure}
	\centering
	\includegraphics[width=\linewidth]{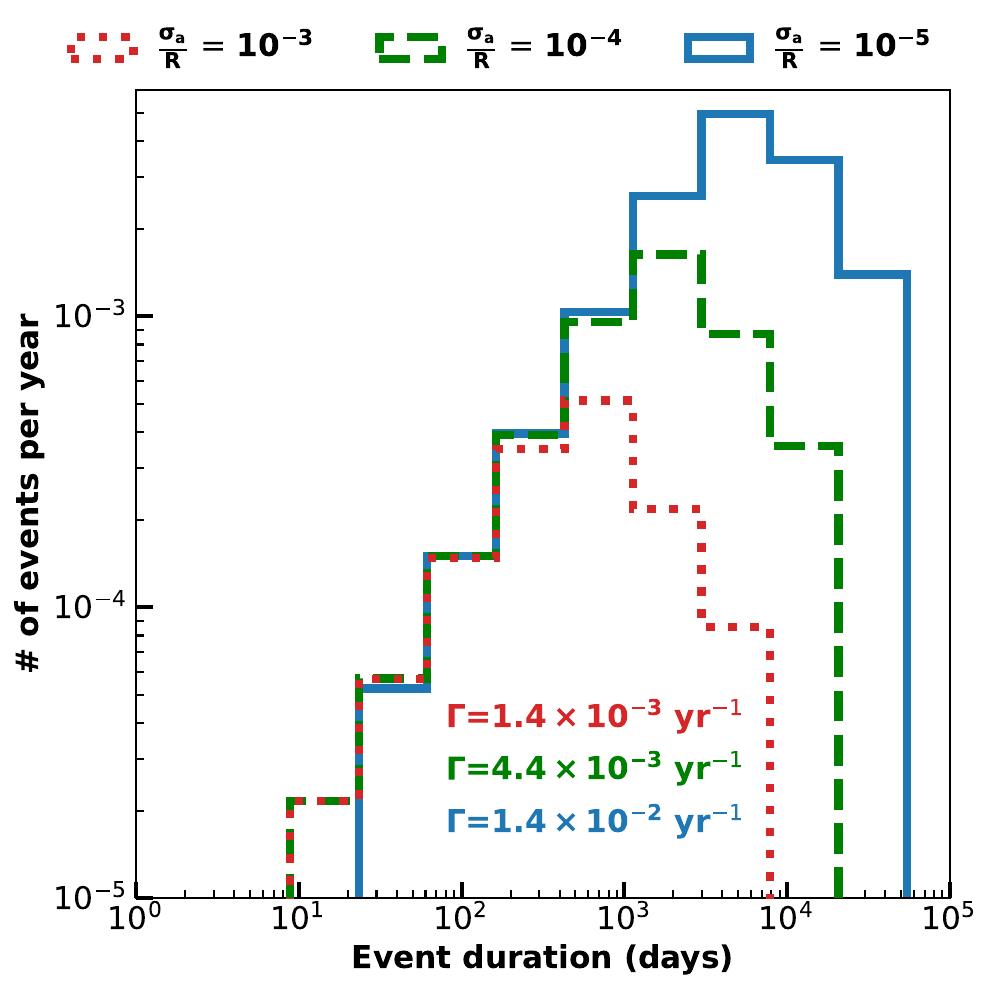}
	\caption{This plot shows the distribution of the event durations $t_e$ of the microlensing events potentially observable in the shadow of Sgr~$^*$. The velocity parameter is assumed to be $100$~km/sec, and the mass $M$ is $1~M_\odot$. The histograms shown with red-dotted, green-dashed, and blue-sold lines correspond to the distributions assuming the uncertainties in the asymmetry observable relative to the radius of the true shadow ($\sigma_a/R$) to be $10^{-3}$, $10^{-4}$, and $10^{-5}$ respectively. These are the typical accuracies achievable with the baseline configurations listed in the table.~\ref{tab:Uncertainty}. The total number of events per year $(\Gamma)$ is also shown for each histogram.}
	\label{fig:EventDurationDist}
\end{figure}

In Figure~\ref{fig:EventDurationDist}, we present the distribution of microlensing event durations based on various accuracies in the asymmetry observable of Sgr~A$^*$. The prevailing trend indicates that a majority of events exhibit durations on the order of $10^2$ days, $10^3$ days, and $10^4$ days, corresponding to $\sigma_a/R$ values of $10^{-3}$, $10^{-4}$, and $10^{-5}$, respectively. Importantly, these durations far exceed the benchmark of one day ($t_\textrm{im}$) and even a month for Earth-space configurations. As a result, the microlensing signatures of the majority of these events remain potentially observable and are not washed out during the imaging time.

Hence, for a given baseline specified by the ratio $\sigma_a/R$, the total event rate can be estimated as
\begin{eqnarray}
    \Gamma &=& \int_{0}^{D_s}\frac{\pi}{2} \xi_\textrm{th}\frac{v}{D_l}\frac{\rho(D_l)}{M}D_l^2dD_l,
\end{eqnarray}
 again assuming the abovementioned stellar density,
\begin{eqnarray}
    \Gamma &\approx1.4\times10^{-3} \textrm{ yr}^{-1}\frac{v}{100~\textrm{km/s}} \left(\frac{D_s}{8.2 \textrm{ kpc}}\right)^{3/2}\sqrt{\frac{1 \textrm{ M}_{\odot}}{M} \frac{10^{-3}}{\sigma_a/R}}.
\end{eqnarray}
The subset of the above events which will induce a detectable magnification in size can then be obtained by dividing the above expression by a constant factor of $38\left(10^{-3}/(\sigma_a/R)\right)^{1/4}$ and given by
\begin{equation}
    \Gamma_\textrm{R$_L$} \approx 3.7\times10^{-5} \textrm{ yr}^{-1}\frac{v}{100~\textrm{km/s}} \sqrt{ \left(\frac{D_s}{8.2 \textrm{ kpc}}\right)^3 \frac{1 \textrm{ M}_{\odot}}{M} \left( \frac{10^{-3}}{\sigma_a/R}\right)^{1/2}}.
\end{equation}

The above expression estimates the number of events potentially observed by baselines at the Moon and $L_2$. These baselines in  space will only have an event rate $\sim$0.0014 event per year due to stellar granularity in the foreground of Sgr~A$^*$. Therefore, the stellar granularity is insufficient to cause a microlensing effect on the shadow of Sgr~A$^*$ very often. However, further investigation is needed with a numerical analysis of the visibility amplitude of the microlensed shadow, to properly model the accuracy of the radial measurements at various baselines. We plan to take this up in our future work. 

\subsection{Potential enhancement in the event rate}
We observe infrequent microlensing events in the black hole shadow of Sgr~A$^*$ caused by conventional stellar lenses toward the Galactic Center, even with sub-micro-arcsec resolution. In this context, we enumerate potential scenarios that could amplify the occurrence of microlensing events in Sgr~A$^*$. Additionally, we outline opportunities to detect microlensing distortions in the black hole shadows located at the cores of various galaxies. This encompasses not only the evident case of M87, where the current EHT has already identified the shadow, but also numerous other shadows anticipated to become accessible through Earth-space interferometry with micro-arcsec or sub-micro-arcsec resolution in the future.
\subsubsection{Black hole cluster around the Galactic Center}
There has been a prediction of a cluster of about 20,000 stellar mass black holes within the central parsec of the Milky Way~\citep{1993ApJ...408..496M, Miralda-Escude:2000kqv,Freitag:2006qf}. This cluster is formed as a result of dynamical friction, which predicts the density profile followed by these black holes will scale as $r^{-7/4}$ and the average mass of these black holes is found to be about $M = 7~M_\odot$~\citep{Miralda-Escude:2000kqv}. Although the cluster of black holes has not been detected yet, a density cusp of quiescent X-ray binaries, which are tracers of isolated black holes, has been detected~\citep{2018Natur.556...70H}. This hints towards the existence of the cluster of black holes within a central parsec of the Milky Way. 
\\
\\
We provide a simple estimate of the contribution in the number of microlensing events from this cluster of black holes, which can potentially be observable in the shadow of Sgr~A$^*$. In general, the density profile of the cluster of $N_\textrm{bh}$ number of black holes within $r_0$ radius from the galactic center can be written as,
\begin{eqnarray}
    \rho(r) &=& \frac{5}{16 \pi} \frac{N_\textrm{bh} M}{r_0^3}\left(\frac{r}{r_0}\right)^{-7/4},
\end{eqnarray}
assuming each black hole has an average mass of $M$. The velocity dispersion of these black holes relative to the galactic center can be obtained using the Jeans equation as~\citep{Chaname:2001an}
\begin{eqnarray}
    v(r) &=& 68.5~\textrm{km/s}\sqrt{\frac{1\textrm{pc}}{r}}.
\end{eqnarray}
\\
\\
\begin{figure}
	\centering
	\includegraphics[width=\linewidth]{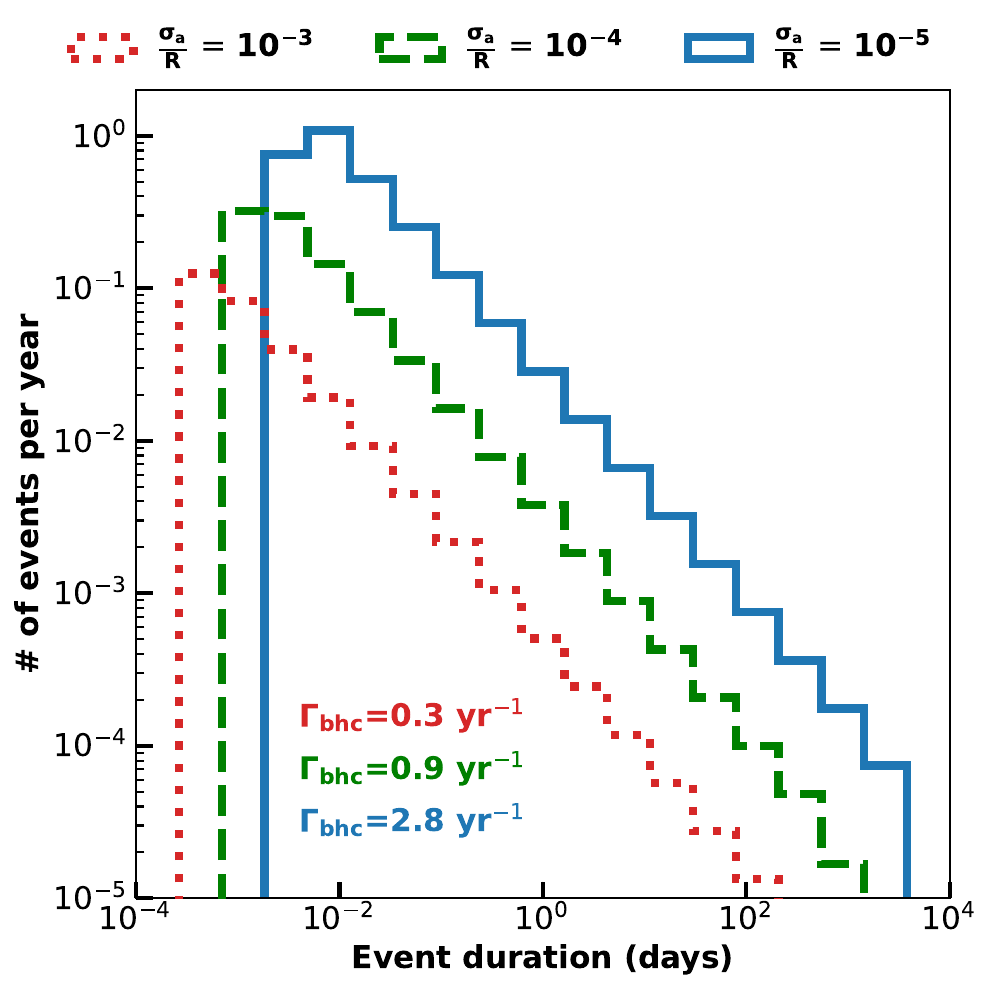}
	\caption{This plot shows the distribution of the event durations $t_e$ of the microlensing events potentially observable in the shadow of Sgr~A$^*$ assuming the black hole mass $M$ to be $7~M_\odot$. The total number of events per year $(\Gamma)$ is shown for each histogram. However,  the event rate of durations above 1~day are $7\times 10^{-4}$, $5\times 10^{-3}$, and $4\times 10^{-2}$ corresponding to $\sigma_a/R$ to be $10^{-3}$, $10^{-4}$, and $10^{-5}$ respectively. These are the typical accuracies achievable with the baseline configurations listed in table.~\ref{tab:Uncertainty}.}
	\label{fig:EventDurationDistBHCluster}
\end{figure}
Following the formalism in the previous sec.~\ref{subsec:EventRate}, we calculate the event distribution  over  event durations under the assumptions  $N_\textrm{bh} = 20,000$ and $r_0 = 1$~pc. The results are shown in fig.~\ref{fig:EventDurationDistBHCluster} under the assumption of various possible values of $\sigma_a/R$ achievable from Earth-space interferometry. We notice the presence of a cluster of black holes can provide a dominant contribution to the microlensing event rate in the shadow. In general, for $N_\textrm{bh}$ number of stellar mass black holes with average mass $M$ within $r_0$ radius from the galactic center, the total number of events can be summarized in the following expression
\begin{equation}
    \Gamma_\textrm{bhc} \approx 0.3 \textrm{ yr}^{-1}\frac{N_\textrm{bh}}{20,000}\left(\frac{1~\textrm{pc}}{r_0}\right)^3 \left(\frac{D_s}{8.2 \textrm{ kpc}}\right)^{3/2}\sqrt{\frac{7 \textrm{ M}_{\odot}}{M} \frac{10^{-3}}{\sigma_a/R}}.
\end{equation}
\\
\\
It is worth noting that the majority of these events exhibit durations below the benchmark imaging time $t_\textrm{im} = 1$ day. Specifically, event rates with durations exceeding one day amount to $7\times 10^{-4}$, $5\times 10^{-3}$, and $4\times 10^{-2}$, corresponding to $\sigma_a/R$ values of $10^{-3}$, $10^{-4}$, and $10^{-5}$ respectively. Thus, detecting these short-duration microlensing events remains challenging due to the potential clustering of black holes around Sgr~A$^*$. However, the considerable number of events with durations smaller than $t_\textrm{im}$ might induce a blurring effect during shadow imaging. This microlensing-induced blurring could potentially be detected by reaching an irreducible angular scale of the order of the Einstein angle. Below this scale, the image may remain unresolved even with the capability of a VLBI facility.
\\
\\
\subsubsection{Other shadows}
We have illustrated the impact of microlensing in the shadow of SgrA$^*$ and examined the potential observability of this phenomenon under various circumstances. A comparable investigation could be conducted for the shadow of M87, evaluating the feasibility of observing such an event by considering a well-informed distribution of compact objects toward M87. Additionally, as black hole imaging achieves higher resolution, numerous other shadows within the capability of EHT-like facilities will be detectable. We anticipate that   sub-millimeter interferometric observations with approximately $0.1\mu$as resolution and $\sim1~\mu$Jy sensitivity may enable the observation of over $\sim 10^6$ supermassive black hole shadows~\citep{Pesce:2021adg}. Consequently, even if the event rates for individual shadows are low, the increased number of measured shadows can enhance the likelihood of observing a microlensed shadow.

\section{Summary}
\label{sec:summary}
We  have presented a formulation of the  phenomenology of microlensing via  observing directly imaged shadows which are potentially detectable in a near-future EHT-like VLBI placed in space. We found an exact analytic expression of the shift in the center and the radial profile of the microlensed shadow. The shift in the center, magnification in the average size, and the asymmetry in shape (defined as the standard deviation in the radial measurements along various position angles) are the key signals imprinted on a microlensed shadow. The maximum average radius of the microlensed shadow could reach up to $3R/2$, and the corresponding maximum asymmetry would be $R/8\sqrt{2}$. 

In the case of the microlensed shadow of Sgr~A$^*$, we find that the maximum average radius is 36.535 $\mu$as, and the maximum asymmetry can reach up to 2.152~$\mu$as. We presented a range of the parameter space of lenses that can be probed with the phenomena of microlensing via future baselines in space. The event rate for future baselines at the Moon and L$_2$ is found to be $1.4\times10^{-3}~\textrm{yr}^{-1}$ for lenses of mass $\sim~M_\odot$ in the stellar component. Therefore, observing the microlensing effect due to such lenses will always be a rare effect. However the signal would be boosted by  an anticipated cluster of stellar mass black holes around the galactic center that would enable us to detect the effect of microlensing in the shadow of Sgr~A$^*$. 

It is worth mentioning here that a more exotic but currently viable scenario of primordial black holes (PBHs) as dark matter \citep{kuhnel} with exciting prospects for future detection via gravitational wave signatures of solar or sub-solar mass candidates \citep{bagui},  should imply an even higher black hole density towards the Galactic Center. This is anticipated in the possible presence of a dark matter spike within the gravitational influence radius ($\sim 0.2~\rm pc$) of  Sgr~A$^*$. This hypothesis is
motivated by theory \citep{gondolo}, constrained by stellar orbits \citep{shen}, and, less directly,  by the observed old stellar density distribution \citep{habibi}.
 Hence the PBH interpretation of dark matter may eventually be tested by estimating the microlensing event rate due to PBHs and comparing it with future high-resolution imaging of the shadow. Additionally, microlensing black hole shadows with the future space-based VLBI facilities will open up exciting possibilities for observing other shadows and hence probing the existence and properties of compact objects in the mass range of $10^{-6} - 10^{4}~M_\odot $ towards the galactic centers.

Lastly, even though we have demonstrated the effect of microlensing on the boundary of the shadow, the formalism presented in this work is equally valid for an impact on the predicted gravitational lensing rings. Microlensing will make the lensing rings larger, asymmetric, and shifted as compared to their true shapes.

\section*{Acknowledgements}

We acknowledge the anonymous reviewer for providing insightful comments on the draft. We would like to thank Vikram Rentala for his valuable suggestions. Author HV would like to acknowledge the hospitality of the \textit{25th Microlensing Conference} held at IAP, Paris, France, and the financial support from IIT Bombay to visit IAP, which gave him an opportunity to start this new collaboration.

\section*{Data Availability}
The data that support the plots shown within this article and other findings of this study will be shared upon reasonable request to the corresponding author.



\bibliographystyle{mnras}
\bibliography{main} 

\begin{thebibliography}{}
\makeatletter
\relax
\def\mn@urlcharsother{\let\do\@makeother \do\$\do\&\do\#\do\^\do\_\do\%\do\~}
\def\mn@doi{\begingroup\mn@urlcharsother \@ifnextchar [ {\mn@doi@}
  {\mn@doi@[]}}
\def\mn@doi@[#1]#2{\def\@tempa{#1}\ifx\@tempa\@empty \href
  {http://dx.doi.org/#2} {doi:#2}\else \href {http://dx.doi.org/#2} {#1}\fi
  \endgroup}
\def\mn@eprint#1#2{\mn@eprint@#1:#2::\@nil}
\def\mn@eprint@arXiv#1{\href {http://arxiv.org/abs/#1} {{\tt arXiv:#1}}}
\def\mn@eprint@dblp#1{\href {http://dblp.uni-trier.de/rec/bibtex/#1.xml}
  {dblp:#1}}
\def\mn@eprint@#1:#2:#3:#4\@nil{\def\@tempa {#1}\def\@tempb {#2}\def\@tempc
  {#3}\ifx \@tempc \@empty \let \@tempc \@tempb \let \@tempb \@tempa \fi \ifx
  \@tempb \@empty \def\@tempb {arXiv}\fi \@ifundefined
  {mn@eprint@\@tempb}{\@tempb:\@tempc}{\expandafter \expandafter \csname
  mn@eprint@\@tempb\endcsname \expandafter{\@tempc}}}

\bibitem[\protect\citeauthoryear{{Abolmasov} \& {Shakura}}{{Abolmasov} \&
  {Shakura}}{2012}]{2012MNRAS.423..676A}
{Abolmasov} P.,  {Shakura} N.~I.,  2012, \mn@doi [\mnras]
  {10.1111/j.1365-2966.2012.20904.x}, \href
  {https://ui.adsabs.harvard.edu/abs/2012MNRAS.423..676A} {423, 676}

\bibitem[\protect\citeauthoryear{{Agol} \& {Krolik}}{{Agol} \&
  {Krolik}}{1999}]{1999ApJ...524...49A}
{Agol} E.,  {Krolik} J.,  1999, \mn@doi [\apj] {10.1086/307800}, \href
  {https://ui.adsabs.harvard.edu/abs/1999ApJ...524...49A} {524, 49}

\bibitem[\protect\citeauthoryear{Akiyama et~al.}{Akiyama
  et~al.}{2019a}]{EventHorizonTelescope:2019dse}
Akiyama K.,  et~al., 2019a, \mn@doi [Astrophys. J. Lett.]
  {10.3847/2041-8213/ab0ec7}, 875, L1

\bibitem[\protect\citeauthoryear{Akiyama et~al.}{Akiyama
  et~al.}{2019b}]{EventHorizonTelescope:2019uob}
Akiyama K.,  et~al., 2019b, \mn@doi [Astrophys. J. Lett.]
  {10.3847/2041-8213/ab0c96}, 875, L2

\bibitem[\protect\citeauthoryear{Akiyama et~al.}{Akiyama
  et~al.}{2019c}]{EventHorizonTelescope:2019jan}
Akiyama K.,  et~al., 2019c, \mn@doi [Astrophys. J. Lett.]
  {10.3847/2041-8213/ab0c57}, 875, L3

\bibitem[\protect\citeauthoryear{Akiyama et~al.}{Akiyama
  et~al.}{2019d}]{EventHorizonTelescope:2019ths}
Akiyama K.,  et~al., 2019d, \mn@doi [Astrophys. J. Lett.]
  {10.3847/2041-8213/ab0e85}, 875, L4

\bibitem[\protect\citeauthoryear{Akiyama et~al.}{Akiyama
  et~al.}{2019e}]{EventHorizonTelescope:2019pgp}
Akiyama K.,  et~al., 2019e, \mn@doi [Astrophys. J. Lett.]
  {10.3847/2041-8213/ab0f43}, 875, L5

\bibitem[\protect\citeauthoryear{Akiyama et~al.}{Akiyama
  et~al.}{2019f}]{EventHorizonTelescope:2019ggy}
Akiyama K.,  et~al., 2019f, \mn@doi [Astrophys. J. Lett.]
  {10.3847/2041-8213/ab1141}, 875, L6

\bibitem[\protect\citeauthoryear{{Bagui} et~al.,}{{Bagui} et~al.}{2023}]{bagui}
{Bagui} E.,  et~al., 2023, \mn@doi [arXiv e-prints]
  {10.48550/arXiv.2310.19857}, \href
  {https://ui.adsabs.harvard.edu/abs/2023arXiv231019857B} {p. arXiv:2310.19857}

\bibitem[\protect\citeauthoryear{Bambi}{Bambi}{2017a}]{Bambi:2015rda}
Bambi C.,  2017a, in {14th Marcel Grossmann Meeting on Recent Developments in
  Theoretical and Experimental General Relativity, Astrophysics, and
  Relativistic Field Theories}. pp 3494--3499 (\mn@eprint {arXiv}
  {1507.05257}), \mn@doi{10.1142/9789813226609_0450}

\bibitem[\protect\citeauthoryear{{Bambi}}{{Bambi}}{2017b}]{2017RvMP...89b5001B}
{Bambi} C.,  2017b, \mn@doi [Reviews of Modern Physics]
  {10.1103/RevModPhys.89.025001}, \href
  {https://ui.adsabs.harvard.edu/abs/2017RvMP...89b5001B} {89, 025001}

\bibitem[\protect\citeauthoryear{Bambi \& Freese}{Bambi \&
  Freese}{2009}]{Bambi:2008jg}
Bambi C.,  Freese K.,  2009, \mn@doi [Phys. Rev. D]
  {10.1103/PhysRevD.79.043002}, 79, 043002

\bibitem[\protect\citeauthoryear{Bambi, Caravelli  \& Modesto}{Bambi
  et~al.}{2012}]{Bambi:2011yz}
Bambi C.,  Caravelli F.,   Modesto L.,  2012, \mn@doi [Phys. Lett. B]
  {10.1016/j.physletb.2012.03.068}, 711, 10

\bibitem[\protect\citeauthoryear{Bambi, Freese, Vagnozzi  \& Visinelli}{Bambi
  et~al.}{2019}]{Bambi:2019tjh}
Bambi C.,  Freese K.,  Vagnozzi S.,   Visinelli L.,  2019, \mn@doi [Phys. Rev.
  D] {10.1103/PhysRevD.100.044057}, 100, 044057

\bibitem[\protect\citeauthoryear{Bardeen}{Bardeen}{1974}]{1974IAUS...64..132B}
Bardeen J.~M.,  1974, \mn@doi [Symposium - International Astronomical Union]
  {10.1017/S0074180900236218}, 64, 132–144

\bibitem[\protect\citeauthoryear{{Baumgardt}, {Amaro-Seoane}  \&
  {Sch{\"o}del}}{{Baumgardt} et~al.}{2018}]{2018A&A...609A..28B}
{Baumgardt} H.,  {Amaro-Seoane} P.,   {Sch{\"o}del} R.,  2018, \mn@doi [\aap]
  {10.1051/0004-6361/201730462}, \href
  {https://ui.adsabs.harvard.edu/abs/2018A&A...609A..28B} {609, A28}

\bibitem[\protect\citeauthoryear{{Boehle} et~al.,}{{Boehle}
  et~al.}{2016}]{2016ApJ...830...17B}
{Boehle} A.,  et~al., 2016, \mn@doi [\apj] {10.3847/0004-637X/830/1/17}, \href
  {https://ui.adsabs.harvard.edu/abs/2016ApJ...830...17B} {830, 17}

\bibitem[\protect\citeauthoryear{{Bozza}}{{Bozza}}{2010}]{2010GReGr..42.2269B}
{Bozza} V.,  2010, \mn@doi [General Relativity and Gravitation]
  {10.1007/s10714-010-0988-2}, \href
  {https://ui.adsabs.harvard.edu/abs/2010GReGr..42.2269B} {42, 2269}

\bibitem[\protect\citeauthoryear{{Bozza}, {de Luca}  \& {Scarpetta}}{{Bozza}
  et~al.}{2006}]{2006PhRvD..74f3001B}
{Bozza} V.,  {de Luca} F.,   {Scarpetta} G.,  2006, \mn@doi [\prd]
  {10.1103/PhysRevD.74.063001}, \href
  {https://ui.adsabs.harvard.edu/abs/2006PhRvD..74f3001B} {74, 063001}

\bibitem[\protect\citeauthoryear{{Carr} \& {Kuhnel}}{{Carr} \&
  {Kuhnel}}{2021}]{kuhnel}
{Carr} B.,  {Kuhnel} F.,  2021, \mn@doi [arXiv e-prints]
  {10.48550/arXiv.2110.02821}, \href
  {https://ui.adsabs.harvard.edu/abs/2021arXiv211002821C} {p. arXiv:2110.02821}

\bibitem[\protect\citeauthoryear{{Chael}, {Issaoun}, {Pesce}, {Johnson},
  {Ricarte}, {Fromm}  \& {Mizuno}}{{Chael} et~al.}{2023}]{Chael:2022meh}
{Chael} A.,  {Issaoun} S.,  {Pesce} D.~W.,  {Johnson} M.~D.,  {Ricarte} A.,
  {Fromm} C.~M.,   {Mizuno} Y.,  2023, \mn@doi [\apj]
  {10.3847/1538-4357/acb7e4}, \href
  {https://ui.adsabs.harvard.edu/abs/2023ApJ...945...40C} {945, 40}

\bibitem[\protect\citeauthoryear{Chaname, Gould  \& Miralda-Escude}{Chaname
  et~al.}{2001}]{Chaname:2001an}
Chaname J.,  Gould A.,   Miralda-Escude J.,  2001, \mn@doi [Astrophys. J.]
  {10.1086/323986}, 563, 793

\bibitem[\protect\citeauthoryear{Cunha \& Herdeiro}{Cunha \&
  Herdeiro}{2018}]{Cunha:2018acu}
Cunha P. V.~P.,  Herdeiro C. A.~R.,  2018, \mn@doi [Gen. Rel. Grav.]
  {10.1007/s10714-018-2361-9}, 50, 42

\bibitem[\protect\citeauthoryear{Cunha, Herdeiro, Radu  \& Runarsson}{Cunha
  et~al.}{2015}]{Cunha:2015yba}
Cunha P. V.~P.,  Herdeiro C. A.~R.,  Radu E.,   Runarsson H.~F.,  2015, \mn@doi
  [Phys. Rev. Lett.] {10.1103/PhysRevLett.115.211102}, 115, 211102

\bibitem[\protect\citeauthoryear{Cunha, Herdeiro  \& Radu}{Cunha
  et~al.}{2019}]{Cunha:2019ikd}
Cunha P. V.~P.,  Herdeiro C. A.~R.,   Radu E.,  2019, \mn@doi [Universe]
  {10.3390/universe5120220}, 5, 220

\bibitem[\protect\citeauthoryear{D'Orazio \& Di~Stefano}{D'Orazio \&
  Di~Stefano}{2018}]{DOrazio:2017ssb}
D'Orazio D.~J.,  Di~Stefano R.,  2018, \mn@doi [Mon. Not. Roy. Astron. Soc.]
  {10.1093/mnras/stx2936}, 474, 2975

\bibitem[\protect\citeauthoryear{D'Orazio \& Di~Stefano}{D'Orazio \&
  Di~Stefano}{2020}]{DOrazio:2019ens}
D'Orazio D.~J.,  Di~Stefano R.,  2020, \mn@doi [Mon. Not. Roy. Astron. Soc.]
  {10.1093/mnras/stz3086}, 491, 1506

\bibitem[\protect\citeauthoryear{Davelaar \& Haiman}{Davelaar \&
  Haiman}{2022a}]{Davelaar:2021gxx}
Davelaar J.,  Haiman Z.,  2022a, \mn@doi [Phys. Rev. D]
  {10.1103/PhysRevD.105.103010}, 105, 103010

\bibitem[\protect\citeauthoryear{Davelaar \& Haiman}{Davelaar \&
  Haiman}{2022b}]{Davelaar:2021eoi}
Davelaar J.,  Haiman Z.,  2022b, \mn@doi [Phys. Rev. Lett.]
  {10.1103/PhysRevLett.128.191101}, 128, 191101

\bibitem[\protect\citeauthoryear{{Deme}, {Meiron}  \& {Kocsis}}{{Deme}
  et~al.}{2020}]{Deme:2019zyv}
{Deme} B.,  {Meiron} Y.,   {Kocsis} B.,  2020, \mn@doi [\apj]
  {10.3847/1538-4357/ab7921}, \href
  {https://ui.adsabs.harvard.edu/abs/2020ApJ...892..130D} {892, 130}

\bibitem[\protect\citeauthoryear{{Do} et~al.,}{{Do}
  et~al.}{2019}]{2019Sci...365..664D}
{Do} T.,  et~al., 2019, \mn@doi [Science] {10.1126/science.aav8137}, \href
  {https://ui.adsabs.harvard.edu/abs/2019Sci...365..664D} {365, 664}

\bibitem[\protect\citeauthoryear{{Dominik} \& {Sahu}}{{Dominik} \&
  {Sahu}}{2000}]{2000ApJ...534..213D}
{Dominik} M.,  {Sahu} K.~C.,  2000, \mn@doi [\apj] {10.1086/308716}, \href
  {https://ui.adsabs.harvard.edu/abs/2000ApJ...534..213D} {534, 213}

\bibitem[\protect\citeauthoryear{{Event Horizon Telescope Collaboration} \&
  {Akiyama}}{{Event Horizon Telescope Collaboration} \&
  {Akiyama}}{2022a}]{2022ApJ...930L..12E}
{Event Horizon Telescope Collaboration} {Akiyama} K. e.~a.,  2022a, \mn@doi
  [\apjl] {10.3847/2041-8213/ac6674}, \href
  {https://ui.adsabs.harvard.edu/abs/2022ApJ...930L..12E} {930, L12}

\bibitem[\protect\citeauthoryear{{Event Horizon Telescope Collaboration} \&
  {Akiyama}}{{Event Horizon Telescope Collaboration} \&
  {Akiyama}}{2022b}]{2022ApJ...930L..13E}
{Event Horizon Telescope Collaboration} {Akiyama} K. a.~a.,  2022b, \mn@doi
  [\apjl] {10.3847/2041-8213/ac6675}, \href
  {https://ui.adsabs.harvard.edu/abs/2022ApJ...930L..13E} {930, L13}

\bibitem[\protect\citeauthoryear{{Event Horizon Telescope Collaboration} \&
  {Akiyama}}{{Event Horizon Telescope Collaboration} \&
  {Akiyama}}{2022c}]{2022ApJ...930L..14E}
{Event Horizon Telescope Collaboration} {Akiyama} K. e.~a.,  2022c, \mn@doi
  [\apjl] {10.3847/2041-8213/ac6429}, \href
  {https://ui.adsabs.harvard.edu/abs/2022ApJ...930L..14E} {930, L14}

\bibitem[\protect\citeauthoryear{Falcke \& Markoff}{Falcke \&
  Markoff}{2013a}]{Falcke:2013ola}
Falcke H.,  Markoff S.~B.,  2013a, \mn@doi [Class. Quant. Grav.]
  {10.1088/0264-9381/30/24/244003}, 30, 244003

\bibitem[\protect\citeauthoryear{{Falcke} \& {Markoff}}{{Falcke} \&
  {Markoff}}{2013b}]{2013CQGra..30x4003F}
{Falcke} H.,  {Markoff} S.~B.,  2013b, \mn@doi [Classical and Quantum Gravity]
  {10.1088/0264-9381/30/24/244003}, \href
  {https://ui.adsabs.harvard.edu/abs/2013CQGra..30x4003F} {30, 244003}

\bibitem[\protect\citeauthoryear{Falcke, Melia  \& Agol}{Falcke
  et~al.}{2000}]{Falcke:1999pj}
Falcke H.,  Melia F.,   Agol E.,  2000, \mn@doi [Astrophys. J. Lett.]
  {10.1086/312423}, 528, L13

\bibitem[\protect\citeauthoryear{{Fish}, {Shea}  \& {Akiyama}}{{Fish}
  et~al.}{2020}]{2020AdSpR..65..821F}
{Fish} V.~L.,  {Shea} M.,   {Akiyama} K.,  2020, \mn@doi [Advances in Space
  Research] {10.1016/j.asr.2019.03.029}, \href
  {https://ui.adsabs.harvard.edu/abs/2020AdSpR..65..821F} {65, 821}

\bibitem[\protect\citeauthoryear{{Fluke} \& {Webster}}{{Fluke} \&
  {Webster}}{1999}]{1999MNRAS.302...68F}
{Fluke} C.~J.,  {Webster} R.~L.,  1999, \mn@doi [\mnras]
  {10.1046/j.1365-8711.1999.02109.x}, \href
  {https://ui.adsabs.harvard.edu/abs/1999MNRAS.302...68F} {302, 68}

\bibitem[\protect\citeauthoryear{Freitag, Amaro-Seoane  \& Kalogera}{Freitag
  et~al.}{2006}]{Freitag:2006qf}
Freitag M.,  Amaro-Seoane P.,   Kalogera V.,  2006, \mn@doi [Astrophys. J.]
  {10.1086/506193}, 649, 91

\bibitem[\protect\citeauthoryear{{Gallego-Cano}, {Sch{\"o}del}, {Dong},
  {Nogueras-Lara}, {Gallego-Calvente}, {Amaro-Seoane}  \&
  {Baumgardt}}{{Gallego-Cano} et~al.}{2018}]{2018A&A...609A..26G}
{Gallego-Cano} E.,  {Sch{\"o}del} R.,  {Dong} H.,  {Nogueras-Lara} F.,
  {Gallego-Calvente} A.~T.,  {Amaro-Seoane} P.,   {Baumgardt} H.,  2018,
  \mn@doi [\aap] {10.1051/0004-6361/201730451}, \href
  {https://ui.adsabs.harvard.edu/abs/2018A&A...609A..26G} {609, A26}

\bibitem[\protect\citeauthoryear{{Gaudi}}{{Gaudi}}{2012}]{2012ARA&A..50..411G}
{Gaudi} B.~S.,  2012, \mn@doi [\araa] {10.1146/annurev-astro-081811-125518},
  \href {https://ui.adsabs.harvard.edu/abs/2012ARA&A..50..411G} {50, 411}

\bibitem[\protect\citeauthoryear{{Gondolo} \& {Silk}}{{Gondolo} \&
  {Silk}}{1999}]{gondolo}
{Gondolo} P.,  {Silk} J.,  1999, \mn@doi [\prl] {10.1103/PhysRevLett.83.1719},
  \href {https://ui.adsabs.harvard.edu/abs/1999PhRvL..83.1719G} {83, 1719}

\bibitem[\protect\citeauthoryear{Gott, Ayzenberg, Yunes  \& Lohfink}{Gott
  et~al.}{2019}]{Gott:2018ocn}
Gott H.,  Ayzenberg D.,  Yunes N.,   Lohfink A.,  2019, \mn@doi [Class. Quant.
  Grav.] {10.1088/1361-6382/ab01b0}, 36, 055007

\bibitem[\protect\citeauthoryear{Gould}{Gould}{1996}]{Gould:1996nb}
Gould A.,  1996, \mn@doi [Publ. Astron. Soc. Pac.] {10.1086/133752}, 108, 465

\bibitem[\protect\citeauthoryear{{Gould} \& {Miralda-Escud{\'e}}}{{Gould} \&
  {Miralda-Escud{\'e}}}{1997}]{1997ApJ...483L..13G}
{Gould} A.,  {Miralda-Escud{\'e}} J.,  1997, \mn@doi [\apjl] {10.1086/310739},
  \href {https://ui.adsabs.harvard.edu/abs/1997ApJ...483L..13G} {483, L13}

\bibitem[\protect\citeauthoryear{Gralla \& Lupsasca}{Gralla \&
  Lupsasca}{2020}]{Gralla:2020yvo}
Gralla S.~E.,  Lupsasca A.,  2020, \mn@doi [Phys. Rev. D]
  {10.1103/PhysRevD.102.124003}, 102, 124003

\bibitem[\protect\citeauthoryear{Gralla, Holz  \& Wald}{Gralla
  et~al.}{2019}]{Gralla:2019xty}
Gralla S.~E.,  Holz D.~E.,   Wald R.~M.,  2019, \mn@doi [Phys. Rev. D]
  {10.1103/PhysRevD.100.024018}, 100, 024018

\bibitem[\protect\citeauthoryear{{Grieger}, {Kayser}  \& {Refsdal}}{{Grieger}
  et~al.}{1988}]{1988A&A...194...54G}
{Grieger} B.,  {Kayser} R.,   {Refsdal} S.,  1988, \aap, \href
  {https://ui.adsabs.harvard.edu/abs/1988A&A...194...54G} {194, 54}

\bibitem[\protect\citeauthoryear{{Gurvits}}{{Gurvits}}{2020}]{2020AdSpR..65..868G}
{Gurvits} L.~I.,  2020, \mn@doi [Advances in Space Research]
  {10.1016/j.asr.2019.05.042}, \href
  {https://ui.adsabs.harvard.edu/abs/2020AdSpR..65..868G} {65, 868}

\bibitem[\protect\citeauthoryear{{Gurvits} et~al.,}{{Gurvits}
  et~al.}{2021}]{2021ExA....51..559G}
{Gurvits} L.~I.,  et~al., 2021, \mn@doi [Experimental Astronomy]
  {10.1007/s10686-021-09714-y}, \href
  {https://ui.adsabs.harvard.edu/abs/2021ExA....51..559G} {51, 559}

\bibitem[\protect\citeauthoryear{Gurvits et~al.}{Gurvits
  et~al.}{2022}]{Gurvits:2022wgm}
Gurvits L.~I.,  et~al., 2022, \mn@doi [Acta Astronaut.]
  {10.1016/j.actaastro.2022.04.020}, 196, 314

\bibitem[\protect\citeauthoryear{{Habibi} et~al.,}{{Habibi}
  et~al.}{2019}]{habibi}
{Habibi} M.,  et~al., 2019, \mn@doi [\apjl] {10.3847/2041-8213/ab03cf}, \href
  {https://ui.adsabs.harvard.edu/abs/2019ApJ...872L..15H} {872, L15}

\bibitem[\protect\citeauthoryear{{Hailey}, {Mori}, {Bauer}, {Berkowitz}, {Hong}
   \& {Hord}}{{Hailey} et~al.}{2018}]{2018Natur.556...70H}
{Hailey} C.~J.,  {Mori} K.,  {Bauer} F.~E.,  {Berkowitz} M.~E.,  {Hong} J.,
  {Hord} B.~J.,  2018, \mn@doi [\nat] {10.1038/nature25029}, \href
  {https://ui.adsabs.harvard.edu/abs/2018Natur.556...70H} {556, 70}

\bibitem[\protect\citeauthoryear{{Hernquist}}{{Hernquist}}{1990}]{1990ApJ...356..359H}
{Hernquist} L.,  1990, \mn@doi [\apj] {10.1086/168845}, \href
  {https://ui.adsabs.harvard.edu/abs/1990ApJ...356..359H} {356, 359}

\bibitem[\protect\citeauthoryear{{Hog}, {Novikov}  \& {Polnarev}}{{Hog}
  et~al.}{1995}]{1995A&A...294..287H}
{Hog} E.,  {Novikov} I.~D.,   {Polnarev} A.~G.,  1995, \aap, \href
  {https://ui.adsabs.harvard.edu/abs/1995A&A...294..287H} {294, 287}

\bibitem[\protect\citeauthoryear{Ingram, Motta, Aigrain  \&
  Karastergiou}{Ingram et~al.}{2021}]{Ingram:2021gar}
Ingram A.,  Motta S.~E.,  Aigrain S.,   Karastergiou A.,  2021, \mn@doi [Mon.
  Not. Roy. Astron. Soc.] {10.1093/mnras/stab609}, 503, 1703

\bibitem[\protect\citeauthoryear{{Irwin}, {Webster}, {Hewett}, {Corrigan}  \&
  {Jedrzejewski}}{{Irwin} et~al.}{1989}]{1989AJ.....98.1989I}
{Irwin} M.~J.,  {Webster} R.~L.,  {Hewett} P.~C.,  {Corrigan} R.~T.,
  {Jedrzejewski} R.~I.,  1989, \mn@doi [\aj] {10.1086/115272}, \href
  {https://ui.adsabs.harvard.edu/abs/1989AJ.....98.1989I} {98, 1989}

\bibitem[\protect\citeauthoryear{Johannsen}{Johannsen}{2016}]{Johannsen:2016uoh}
Johannsen T.,  2016, \mn@doi [Class. Quant. Grav.]
  {10.1088/0264-9381/33/12/124001}, 33, 124001

\bibitem[\protect\citeauthoryear{Johannsen \& Psaltis}{Johannsen \&
  Psaltis}{2010a}]{Johannsen:2010xs}
Johannsen T.,  Psaltis D.,  2010a, \mn@doi [Astrophys. J.]
  {10.1088/0004-637X/716/1/187}, 716, 187

\bibitem[\protect\citeauthoryear{Johannsen \& Psaltis}{Johannsen \&
  Psaltis}{2010b}]{Johannsen:2010ru}
Johannsen T.,  Psaltis D.,  2010b, \mn@doi [Astrophys. J.]
  {10.1088/0004-637X/718/1/446}, 718, 446

\bibitem[\protect\citeauthoryear{{Johannsen} \& {Psaltis}}{{Johannsen} \&
  {Psaltis}}{2010c}]{2010ApJ...718..446J}
{Johannsen} T.,  {Psaltis} D.,  2010c, \mn@doi [\apj]
  {10.1088/0004-637X/718/1/446}, \href
  {https://ui.adsabs.harvard.edu/abs/2010ApJ...718..446J} {718, 446}

\bibitem[\protect\citeauthoryear{Johnson et~al.}{Johnson
  et~al.}{2020}]{Johnson:2019ljv}
Johnson M.~D.,  et~al., 2020, \mn@doi [Sci. Adv.] {10.1126/sciadv.aaz1310}, 6,
  eaaz1310

\bibitem[\protect\citeauthoryear{{Kayser}, {Refsdal}  \& {Stabell}}{{Kayser}
  et~al.}{1986}]{1986A&A...166...36K}
{Kayser} R.,  {Refsdal} S.,   {Stabell} R.,  1986, \aap, \href
  {https://ui.adsabs.harvard.edu/abs/1986A&A...166...36K} {166, 36}

\bibitem[\protect\citeauthoryear{{Kiraga} \& {Paczynski}}{{Kiraga} \&
  {Paczynski}}{1994}]{1994ApJ...430L.101K}
{Kiraga} M.,  {Paczynski} B.,  1994, \mn@doi [\apjl] {10.1086/187448}, \href
  {https://ui.adsabs.harvard.edu/abs/1994ApJ...430L.101K} {430, L101}

\bibitem[\protect\citeauthoryear{{Kochanek}}{{Kochanek}}{2004}]{2004ApJ...605...58K}
{Kochanek} C.~S.,  2004, \mn@doi [\apj] {10.1086/382180}, \href
  {https://ui.adsabs.harvard.edu/abs/2004ApJ...605...58K} {605, 58}

\bibitem[\protect\citeauthoryear{{Li}}{{Li}}{2016}]{2016arXiv161207781L}
{Li} E.,  2016, \mn@doi [arXiv e-prints] {10.48550/arXiv.1612.07781}, \href
  {https://ui.adsabs.harvard.edu/abs/2016arXiv161207781L} {p. arXiv:1612.07781}

\bibitem[\protect\citeauthoryear{{Likhachev}, {Rudnitskiy}, {Shchurov},
  {Andrianov}, {Baryshev}, {Chernov}  \& {Kostenko}}{{Likhachev}
  et~al.}{2022}]{2022MNRAS.511..668L}
{Likhachev} S.~F.,  {Rudnitskiy} A.~G.,  {Shchurov} M.~A.,  {Andrianov} A.~S.,
  {Baryshev} A.~M.,  {Chernov} S.~V.,   {Kostenko} V.~I.,  2022, \mn@doi
  [\mnras] {10.1093/mnras/stac079}, \href
  {https://ui.adsabs.harvard.edu/abs/2022MNRAS.511..668L} {511, 668}

\bibitem[\protect\citeauthoryear{{Lindegren} et~al.,}{{Lindegren}
  et~al.}{2021}]{2021A&A...649A...2L}
{Lindegren} L.,  et~al., 2021, \mn@doi [\aap] {10.1051/0004-6361/202039709},
  \href {https://ui.adsabs.harvard.edu/abs/2021A&A...649A...2L} {649, A2}

\bibitem[\protect\citeauthoryear{Liu, Eatough, Wex  \& Kramer}{Liu
  et~al.}{2014}]{Liu:2014uka}
Liu K.,  Eatough R.~P.,  Wex N.,   Kramer M.,  2014, \mn@doi [Mon. Not. Roy.
  Astron. Soc.] {10.1093/mnras/stu1913}, 445, 3115

\bibitem[\protect\citeauthoryear{{Luminet}}{{Luminet}}{1979}]{1979A&A....75..228L}
{Luminet} J.~P.,  1979, \aap, \href
  {https://ui.adsabs.harvard.edu/abs/1979A&A....75..228L} {75, 228}

\bibitem[\protect\citeauthoryear{{Mediavilla}, {Jim{\'e}nez-vicente},
  {Mu{\~n}oz}  \& {Mediavilla}}{{Mediavilla}
  et~al.}{2015}]{2015ApJ...814L..26M}
{Mediavilla} E.,  {Jim{\'e}nez-vicente} J.,  {Mu{\~n}oz} J.~A.,   {Mediavilla}
  T.,  2015, \mn@doi [\apjl] {10.1088/2041-8205/814/2/L26}, \href
  {https://ui.adsabs.harvard.edu/abs/2015ApJ...814L..26M} {814, L26}

\bibitem[\protect\citeauthoryear{Mikheeva, Repin  \& Lukash}{Mikheeva
  et~al.}{2020}]{Mikheeva:2020bqj}
Mikheeva E.~V.,  Repin S.~V.,   Lukash V.~N.,  2020, \mn@doi [Astron. Rep.]
  {10.1134/S1063772920080065}, 64, 578

\bibitem[\protect\citeauthoryear{Miralda-Escude \& Gould}{Miralda-Escude \&
  Gould}{2000}]{Miralda-Escude:2000kqv}
Miralda-Escude J.,  Gould A.,  2000, \mn@doi [Astrophys. J.] {10.1086/317837},
  545, 847

\bibitem[\protect\citeauthoryear{{Morris}}{{Morris}}{1993}]{1993ApJ...408..496M}
{Morris} M.,  1993, \mn@doi [\apj] {10.1086/172607}, \href
  {https://ui.adsabs.harvard.edu/abs/1993ApJ...408..496M} {408, 496}

\bibitem[\protect\citeauthoryear{{Mortonson}, {Schechter}  \&
  {Wambsganss}}{{Mortonson} et~al.}{2005}]{2005ApJ...628..594M}
{Mortonson} M.~J.,  {Schechter} P.~L.,   {Wambsganss} J.,  2005, \mn@doi [\apj]
  {10.1086/431195}, \href
  {https://ui.adsabs.harvard.edu/abs/2005ApJ...628..594M} {628, 594}

\bibitem[\protect\citeauthoryear{Narayan \& Bartelmann}{Narayan \&
  Bartelmann}{1996}]{Narayan:1996ba}
Narayan R.,  Bartelmann M.,  1996, in {13th Jerusalem Winter School in
  Theoretical Physics: Formation of Structure in the Universe}.  (\mn@eprint
  {arXiv} {astro-ph/9606001})

\bibitem[\protect\citeauthoryear{{Nucita}, {de Paolis}, {Ingrosso}, {Giordano}
  \& {Manni}}{{Nucita} et~al.}{2017}]{2017IJMPD..2641015N}
{Nucita} A.~A.,  {de Paolis} F.,  {Ingrosso} G.,  {Giordano} M.,   {Manni} L.,
  2017, \mn@doi [International Journal of Modern Physics D]
  {10.1142/S0218271817410152}, \href
  {https://ui.adsabs.harvard.edu/abs/2017IJMPD..2641015N} {26, 1741015}

\bibitem[\protect\citeauthoryear{Paczynski}{Paczynski}{1996}]{Paczynski:1996nh}
Paczynski B.,  1996, \mn@doi [Ann. Rev. Astron. Astrophys.]
  {10.1146/annurev.astro.34.1.419}, 34, 419

\bibitem[\protect\citeauthoryear{{Paugnat}, {Lupsasca}, {Vincent}  \&
  {Wielgus}}{{Paugnat} et~al.}{2022}]{2022A&A...668A..11P}
{Paugnat} H.,  {Lupsasca} A.,  {Vincent} F.~H.,   {Wielgus} M.,  2022, \mn@doi
  [\aap] {10.1051/0004-6361/202244216}, \href
  {https://ui.adsabs.harvard.edu/abs/2022A&A...668A..11P} {668, A11}

\bibitem[\protect\citeauthoryear{Perlick \& Tsupko}{Perlick \&
  Tsupko}{2022}]{Perlick:2021aok}
Perlick V.,  Tsupko O.~Y.,  2022, \mn@doi [Phys. Rept.]
  {10.1016/j.physrep.2021.10.004}, 947, 1

\bibitem[\protect\citeauthoryear{Pesce et~al.,}{Pesce
  et~al.}{2021}]{Pesce:2021adg}
Pesce D.~W.,  et~al., 2021, \mn@doi [Astrophys. J.] {10.3847/1538-4357/ac2eb5},
  923, 260

\bibitem[\protect\citeauthoryear{{Poindexter}, {Morgan}  \&
  {Kochanek}}{{Poindexter} et~al.}{2008}]{2008ApJ...673...34P}
{Poindexter} S.,  {Morgan} N.,   {Kochanek} C.~S.,  2008, \mn@doi [\apj]
  {10.1086/524190}, \href
  {https://ui.adsabs.harvard.edu/abs/2008ApJ...673...34P} {673, 34}

\bibitem[\protect\citeauthoryear{Psaltis, Ozel, Chan  \& Marrone}{Psaltis
  et~al.}{2015}]{Psaltis:2014mca}
Psaltis D.,  Ozel F.,  Chan C.-K.,   Marrone D.~P.,  2015, \mn@doi [Astrophys.
  J.] {10.1088/0004-637X/814/2/115}, 814, 115

\bibitem[\protect\citeauthoryear{{Refsdal}}{{Refsdal}}{1964}]{1964MNRAS.128..295R}
{Refsdal} S.,  1964, \mn@doi [\mnras] {10.1093/mnras/128.4.295}, \href
  {https://ui.adsabs.harvard.edu/abs/1964MNRAS.128..295R} {128, 295}

\bibitem[\protect\citeauthoryear{Roelofs et~al.}{Roelofs
  et~al.}{2019a}]{Roelofs:2019nmh}
Roelofs F.,  et~al., 2019a, \mn@doi [Astron. Astrophys.]
  {10.1051/0004-6361/201732423}, 625, A124

\bibitem[\protect\citeauthoryear{{Roelofs} et~al.,}{{Roelofs}
  et~al.}{2019b}]{2019A&A...625A.124R}
{Roelofs} F.,  et~al., 2019b, \mn@doi [\aap] {10.1051/0004-6361/201732423},
  \href {https://ui.adsabs.harvard.edu/abs/2019A&A...625A.124R} {625, A124}

\bibitem[\protect\citeauthoryear{{Rudnitskiy}, {Mzhelskiy}, {Shchurov},
  {Syachina}  \& {Zapevalin}}{{Rudnitskiy} et~al.}{2022}]{2022AcAau.196...29R}
{Rudnitskiy} A.~G.,  {Mzhelskiy} P.~V.,  {Shchurov} M.~A.,  {Syachina} T.~A.,
  {Zapevalin} P.~R.,  2022, \mn@doi [Acta Astronautica]
  {10.1016/j.actaastro.2022.03.036}, \href
  {https://ui.adsabs.harvard.edu/abs/2022AcAau.196...29R} {196, 29}

\bibitem[\protect\citeauthoryear{{Schmidt} \& {Wambsganss}}{{Schmidt} \&
  {Wambsganss}}{2010}]{2010GReGr..42.2127S}
{Schmidt} R.~W.,  {Wambsganss} J.,  2010, \mn@doi [General Relativity and
  Gravitation] {10.1007/s10714-010-0956-x}, \href
  {https://ui.adsabs.harvard.edu/abs/2010GReGr..42.2127S} {42, 2127}

\bibitem[\protect\citeauthoryear{{Schneider} \& {Weiss}}{{Schneider} \&
  {Weiss}}{1987}]{1987A&A...171...49S}
{Schneider} P.,  {Weiss} A.,  1987, \aap, \href
  {https://ui.adsabs.harvard.edu/abs/1987A&A...171...49S} {171, 49}

\bibitem[\protect\citeauthoryear{{Schneider}, {Ehlers}  \& {Falco}}{{Schneider}
  et~al.}{1992}]{1992grle.book.....S}
{Schneider} P.,  {Ehlers} J.,   {Falco} E.~E.,  1992, {Gravitational Lenses}.
Gravitational Lenses, XIV, 560 pp. 112 figs.. Springer-Verlag Berlin Heidelberg
  New York. Also Astronomy and Astrophysics Library,
  \mn@doi{10.1007/978-3-662-03758-4}

\bibitem[\protect\citeauthoryear{{Sch{\"o}del}, {Gallego-Cano}, {Dong},
  {Nogueras-Lara}, {Gallego-Calvente}, {Amaro-Seoane}  \&
  {Baumgardt}}{{Sch{\"o}del} et~al.}{2018}]{2018A&A...609A..27S}
{Sch{\"o}del} R.,  {Gallego-Cano} E.,  {Dong} H.,  {Nogueras-Lara} F.,
  {Gallego-Calvente} A.~T.,  {Amaro-Seoane} P.,   {Baumgardt} H.,  2018,
  \mn@doi [\aap] {10.1051/0004-6361/201730452}, \href
  {https://ui.adsabs.harvard.edu/abs/2018A&A...609A..27S} {609, A27}

\bibitem[\protect\citeauthoryear{{Shen}, {Yuan}, {Jiang}, {Tsai}, {Yuan}  \&
  {Fan}}{{Shen} et~al.}{2024}]{shen}
{Shen} Z.-Q.,  {Yuan} G.-W.,  {Jiang} C.-Z.,  {Tsai} Y.-L.~S.,  {Yuan} Q.,
  {Fan} Y.-Z.,  2024, \mn@doi [\mnras] {10.1093/mnras/stad3282}, \href
  {https://ui.adsabs.harvard.edu/abs/2024MNRAS.527.3196S} {527, 3196}

\bibitem[\protect\citeauthoryear{{Takahashi}}{{Takahashi}}{2004}]{2004ApJ...611..996T}
{Takahashi} R.,  2004, \mn@doi [\apj] {10.1086/422403}, \href
  {https://ui.adsabs.harvard.edu/abs/2004ApJ...611..996T} {611, 996}

\bibitem[\protect\citeauthoryear{{Tiede}, {Johnson}, {Pesce}, {Palumbo},
  {Chang}  \& {Galison}}{{Tiede} et~al.}{2022}]{Tiede:2022grp}
{Tiede} P.,  {Johnson} M.~D.,  {Pesce} D.~W.,  {Palumbo} D. C.~M.,  {Chang}
  D.~O.,   {Galison} P.,  2022, \mn@doi [Galaxies] {10.3390/galaxies10060111},
  \href {https://ui.adsabs.harvard.edu/abs/2022Galax..10..111T} {10, 111}

\bibitem[\protect\citeauthoryear{{Valenti} et~al.,}{{Valenti}
  et~al.}{2016}]{2016A&A...587L...6V}
{Valenti} E.,  et~al., 2016, \mn@doi [\aap] {10.1051/0004-6361/201527500},
  \href {https://ui.adsabs.harvard.edu/abs/2016A&A...587L...6V} {587, L6}

\bibitem[\protect\citeauthoryear{{Walker}}{{Walker}}{1995}]{1995ApJ...453...37W}
{Walker} M.~A.,  1995, \mn@doi [\apj] {10.1086/176367}, \href
  {https://ui.adsabs.harvard.edu/abs/1995ApJ...453...37W} {453, 37}

\bibitem[\protect\citeauthoryear{Wambsganss}{Wambsganss}{1998}]{Wambsganss:1998gg}
Wambsganss J.,  1998, \mn@doi [Living Rev. Rel.] {10.12942/lrr-1998-12}, 1, 12

\bibitem[\protect\citeauthoryear{{Zhu}, {Johnson}  \& {Narayan}}{{Zhu}
  et~al.}{2019}]{2019ApJ...870....6Z}
{Zhu} Z.,  {Johnson} M.~D.,   {Narayan} R.,  2019, \mn@doi [\apj]
  {10.3847/1538-4357/aaef3d}, \href
  {https://ui.adsabs.harvard.edu/abs/2019ApJ...870....6Z} {870, 6}

\bibitem[\protect\citeauthoryear{{de Vries}}{{de
  Vries}}{2000}]{2000CQGra..17..123D}
{de Vries} A.,  2000, \mn@doi [Classical and Quantum Gravity]
  {10.1088/0264-9381/17/1/309}, \href
  {https://ui.adsabs.harvard.edu/abs/2000CQGra..17..123D} {17, 123}

\makeatother
\end{thebibliography}




\appendix
\section{Other observable plots}
\begin{figure*}
	\centering
	\includegraphics[width=\textwidth]{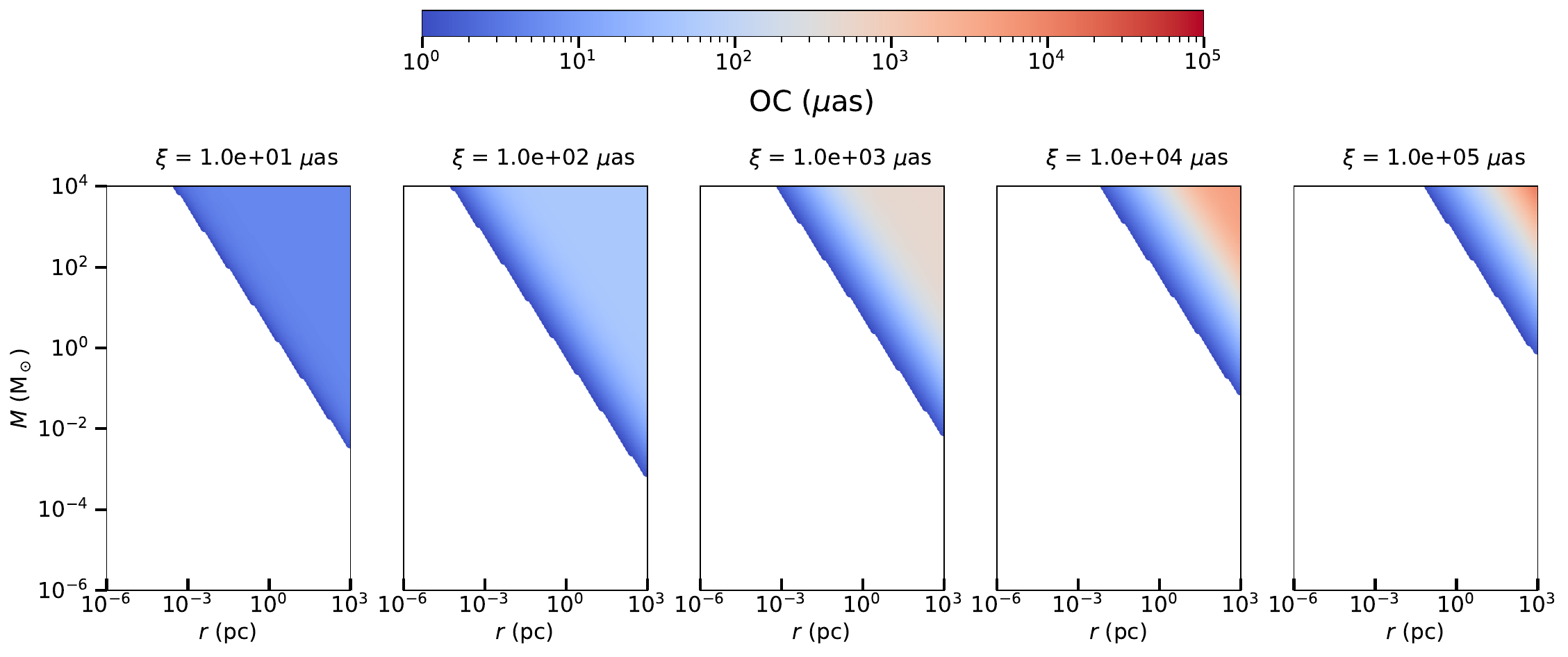}
	\caption{This figure depicts the lens's mass and distance from the center of Sgr~A$^*$ and the corresponding induced shift in the lensed-center. Various panels correspond to the different angular separation of the lens from the shadow of Sgr~A$^*$.}
	\label{fig:ObservableLcenter}
\end{figure*}

We show the induced shift in the lensed center for  different angular separations of the lens from the shadow of Sgr~A$^*$    (fig.~\ref{fig:ObservableLcenter}) as a function of
the lens's mass and distance from the center of Sgr~A$^*$. 

\begin{figure*}
	\centering
	\includegraphics[width=\textwidth]{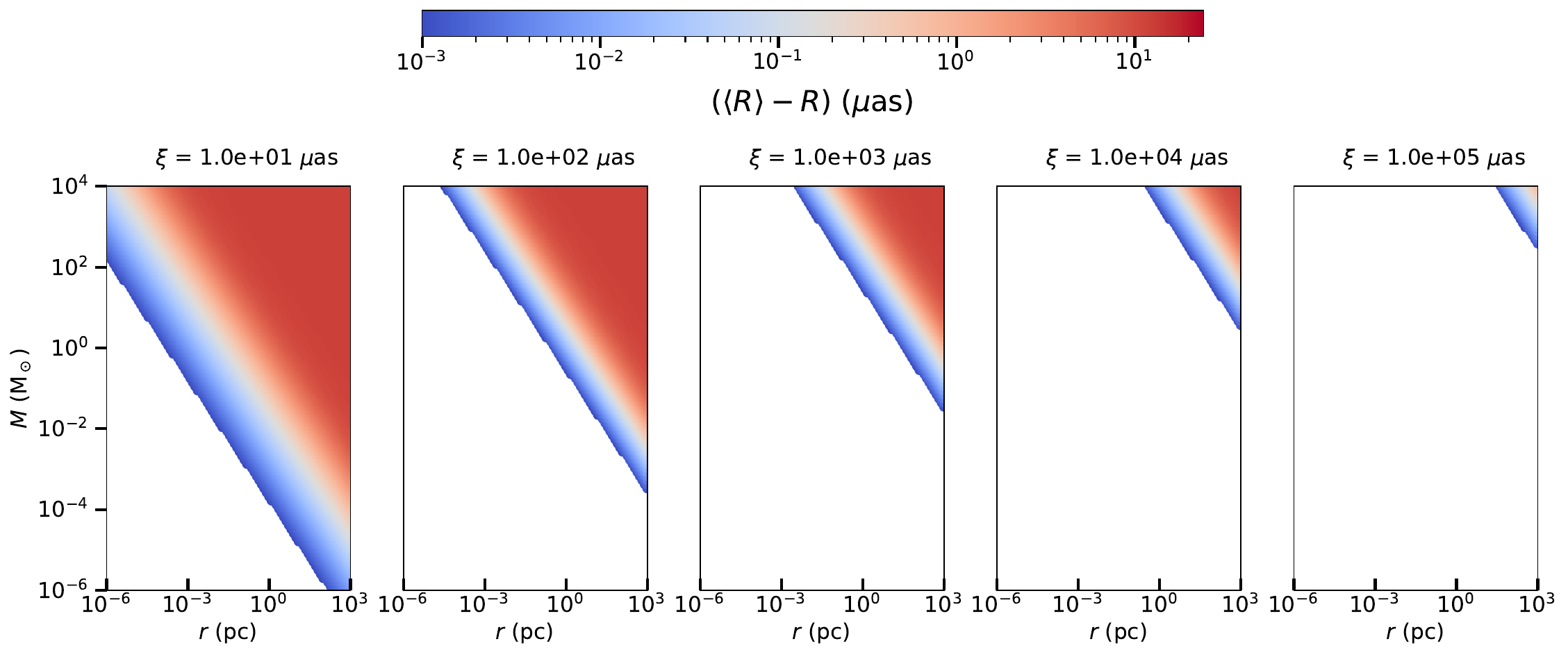}
	\caption{This figure depicts the lens's mass and distance from the center of Sgr~A$^*$ and the corresponding induced magnification in the average radius of the microlensed shadow. Various panels correspond to the different angular separation of the lens from the shadow of Sgr~A$^*$.}
	\label{fig:ObservableRadius}
\end{figure*}

The induced magnification is derived in fig.~\ref{fig:ObservableRadius} for  the average radius of the microlensed shadow and different angular separations of the lens from the center of Sgr~A$^*$. 

\begin{figure*}
	\centering
	\includegraphics[width=\textwidth]{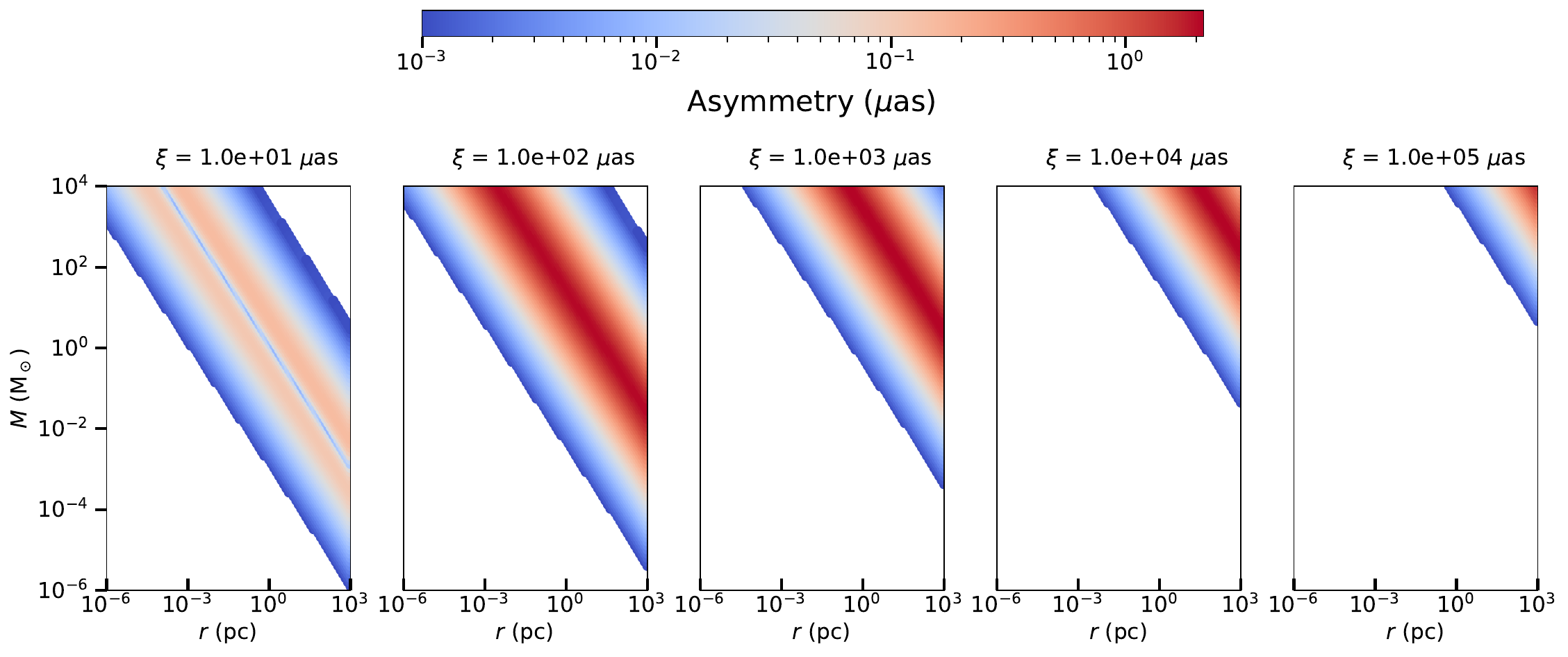}
	\caption{This figure depicts the lens's mass and distance from the center of Sgr~A$^*$ and the corresponding induced asymmetry in the microlensed shadow. Various panels correspond to the different angular separation of the lens from the shadow of Sgr~A$^*$.}
	\label{fig:ObservableAsym}
\end{figure*}

Finally, we depict the induced asymmetry in the microlensed shadow corresponding to different angular separations of lenses of different masses from the  center of Sgr~A$^*$ (fig.~\ref{fig:ObservableAsym}).

\bsp	
\label{lastpage}
\end{document}